\documentstyle[prb,preprint,aps]{revtex}

\begin{document}

\title{He Scattering from Random Adsorbates, Disordered Compact Islands and
Fractal Submonolayers: Intensity Manifestations of Surface Disorder
}

\author{A.T. Yinnon,$^a$ D.A. Lidar (Hamburger),$^{a,b}$ I. Farbman,$^{a}$
R.B. Gerber,$^{a,c}$ P. Zeppenfeld,$^d$ M.A. Krzyzowski,$^d$ and G. Comsa$^d$}
\address{}
\author{}
\address{
$^a$Department of Physical Chemistry and The Fritz Haber Center for Molecular
Dynamics, The Hebrew University of Jerusalem, Jerusalem 91904, Israel\\
$^b$Department of Physics, The Hebrew University of Jerusalem, Jerusalem
91904, Israel\\
$^c$Department of Chemistry, University of California - Irvine, Irvine, CA
92717, USA\\
$^d$Institut f\"{u}r Grenzfl\"{a}chenforschung und Vakuumphysik,
Forschunszentrum J\"{u}lich, Postfach 1913, D-52425 J\"{u}lich, Germany
}

\maketitle

\begin{abstract}
A theoretical study is made on He scattering from three basic classes of
disordered adlayers: (a) Translationally random adsorbates, (b) disordered
compact islands and (c) fractal submonolayers. The implications of the results
to experimental studies of He scattering from disordered surfaces are
discussed, and a combined experimental-theoretical study is made for Ag
submonolayers on Pt(111). Some of the main theoretical findings are: (1) The
scattering intensities from the three disorder classes differ significantly,
and can be used to distinguish between them. (2) Structural aspects of the
calculated intensities from translationally random clusters were found to be
strongly correlated with those of individual clusters. (3) For fractal
islands, just as for all surfaces considered here, the off-specular
intensity depends on the parameters of the He/Ag interaction, and does not
follow a universal power law as previously proposed in the literature.

In the experimental-theoretical study of Ag on Pt(111), we use experimental He
scattering data from low-coverage (single adsorbate) systems to determine an
empirical He/Ag-Pt potential of good quality. Then, we carry out He scattering
calculations for high coverage and compare with experiments for these
systems. The conclusion is that the actual experimental phase corresponds to
small compact Ag clusters of narrow size distribution, with partial
translational disorder.

\end{abstract}

\newpage

\section{Introduction}
\label{introduction}

The properties of thin metal or semiconductor films on solid substrates are of
major theoretical, experimental and technological interest. The structure of
these films influences their physical and chemical properties, and these may be
of central importance, for example, in the fabrication of electronic
devices. The shape of adlayers generally depends on the growth kinetics and
the microscopic characteristics of the substrate. Thin films are typically
produced by epitaxial growth processes, wherein vapor atoms or clusters are
deposited on a substrate.\cite{1,2a,2b,3,4,5a,5b,6} Once adsorbed, the
adatoms diffuse on the substrate surface, and when reaching short mutual
distance they can form stable nuclei, which subsequently grow to clusters or
islands by attachment of further adatoms. Nucleation and growth are competing
processes, which depend on diffusion of adatoms, stability of the adsorbed
clusters, and surface diffusion of these clusters. Therefore, thin film
structures contain a wealth of information about microscopic growth processes,
and adatom/substrate and adatom/adatom interactions.

Experimental studies have shown that in addition to ordered adlayers, various
disordered structures may form. For example on adsorption of Ag on Pt(111)
surfaces, at low substrate temperatures and low deposition rates, isolated
small clusters may form.\cite{2a,2b,3,4} Elevation of the surface temperature
or lowering of the deposition rate can lead to the growth of fractal
islands.\cite{2a,2b} At even higher temperatures randomly distributed compact
islands may form.\cite{2a,2b,3,4} Therefore, thin
films offer an exceptional opportunity for studying two dimensional disordered
systems. By investigating the relation between the structure of thin films and
their growth processes, valuable information can be obtained about the
processes leading to the emergence of different kinds of disorder. This,
however, requires techniques for determining the disordered adlayer
structure.

Surface morphologies can be probed by scanning tunneling microscopy (STM) or
diffraction techniques, such as thermal energy atom scattering (TEAS),
low-energy or reflective high energy electron diffraction (LEED or
RHEED). Diffraction methods have the property that in addition to being
sensitive to local surface features, they resolve the global surface
structure. Accordingly, they provide a better average of the over-all
topography than direct imaging techniques, such as STM. Moreover, diffraction
readily allows for monitoring adlayer growth in situ and at various
temperatures and therefore can easily follow its temporal evolution. For
studying surface structure, a technique of great power is Helium atom
scattering. It is non-destructive, and He only probes the outer layer. Also,
thermal He atom scattering is dominantly elastic. Moreover, since its
wavelength is of the order of surface unit cells, it is very sensitive to the
local adlayer structure. He atom scattering has successfully been employed in
the study of ordered surfaces and surfaces with isolated defects. However,
relatively little is known as yet on the manifestations of different kinds of
disorder in He scattering patterns. Two of the few experimentally and
theoretically studied disordered surfaces are those of substitutionally
disordered mixed Xe+Kr monolayers on Pt(111),\cite{7a} and translationally
disordered small Pt-clusters on Pt(111).\cite{7b} It was found that
attenuation of the specular peak in He scattering from the surface due to the
presence of adsorbates or other defects contains a wealth of information about
adatom/adatom interactions, the clustering of adatoms, and 2D vs. 3D epitaxial
growth.\cite{7b} Also, non-Bragg maxima, e.g., Fraunhofer or rainbow maxima,
appear for such systems. These maxima contain information on the
microstructure of islands and are sensitive to percolation
transitions. Non-specular Bragg peaks appear as well, which contain
information on the corrugation of the He/adlayer interaction.\cite{7a} The
study of randomly adsorbed rare-gas overlayers has the advantage that the
interactions between the rare gas atoms and He are well known. In contrast
little is known about the interactions in a He/(metal adsorbate)/(metal
substrate) system. So far, this has complicated the interpretation of features
in the angular scattering patterns, and has made identifying different kinds
of disorder difficult.

In this paper, for the first time a comparative He scattering study is
undertaken among three major classes of disorder: Translationally random
adatoms, translationally random compact islands, and fractal/dendritic
submonolayers. These classes are representative of a large proportion of the
experimentally reported adlayer structures at submonolayer coverage.\cite{7c}
It is thus of fundamental interest to understand the differences in their
respective He scattering intensity distributions, and to investigate whether
the type of disorder can be identified by a He scattering experiment. In order
to carry out this program reliably, we determined the interaction potential
between He and a Pt(111) surface with an adsorbed Ag atom. From this we have
constructed what we hope is a realistic potential function for He interaction
with a {\em disordered} metal submonolayer. The choice of the Ag/Pt(111)
system was motivated by the fact that the He/Pt(111) potential is rather
smooth parallel to the surface. Therefore the structures in the scattering
patterns are mainly due to the Ag adsorbates. Moreover, the He/Pt(111) belongs
to one of the few atom/surface systems for which a fairly reliable empirical
potential is available.

The outline of the paper is as follows: Section \ref{methods} describes the
experimental and theoretical methods employed, and states the assumptions made
in the theoretical part of this work. Section \ref{results} discusses the
scattering from the different disorder classes. In section
\ref{experiment-theory} we analyze the experimental scattering data and
suggest a morphological identification of the disordered phase present on the
surface. Concluding remarks are presented in section \ref{conclusions}.

\section{Systems and Methods}
\label{methods}

The adlayer structures studied here theoretically are:

\begin{itemize}
\begin{enumerate}
\item{Full Ag monolayer.}
\label{monolayer}
\item{Single adsorbed Ag atom.}
\label{single}
\item{Isolated small compact cluster (henceforth SCC).}
\label{heptamer}
\item{Ag adatoms randomly adsorbed on a lattice.}
\label{lattice-random}
\item{SCCs of Ag randomly adsorbed on a lattice.}
\label{random-heptamers}
\item{Ag adatoms completely randomly and continuously distributed (Ag
adatoms can be arbitrarily close).}
\label{fully-random}
\item{Large compact Ag islands.}
\label{compact}
\item{Fractal islands of Ag atoms.}
\label{fractal}
\end{enumerate}
\end{itemize}

(\ref{monolayer}) and (\ref{single}) are important as a reference, and for
developing potentials. All other cases were chosen as representing important
types of disorder.  In all cases, the adsorbates are on a Pt(111) surface. In
the case of randomly adsorbed Ag adatoms one would expect to observe system
(\ref{lattice-random}) and not the idealized system
(\ref{fully-random}). However, calculations for the latter structure leads to
additional insight into angular scattering patterns.

\subsection{Experimental Methods}
\label{experiment}

The experiments were performed in a high resolution UHV helium scattering
apparatus with a nominal base pressure $<10^{-10}$ mbar. The system is
equipped with a commercial Knudsen cell by means of which high purity
(99.999\%) silver can be evaporated onto the Pt(111) sample at a rate between
$10^{-1}$ and $10^{-4}$ monolayers per second. The Pt(111) sample was cleaned
by repeated cycles of heating in an oxygen atmosphere ($10^{-6}$ mbar) at
750K, sputtering with 1 keV Ar-ions followed by short annealing at 1250K. The
sample is mounted on a manipulator which allows the crystal to be rotated
around its three principal axes. In addition, the total angle $\chi = \theta_i
+ \theta_f$ of the incident and outgoing He-beam can be varied between
60$^\circ$ and 120$^\circ$ by rotating the detector. By means of a liquid
helium cryostat the sample can be cooled to 20K and by concomitant heating by
electron bombardment the surface temperature can be set and held constant at
any value between 20K and 1250K.

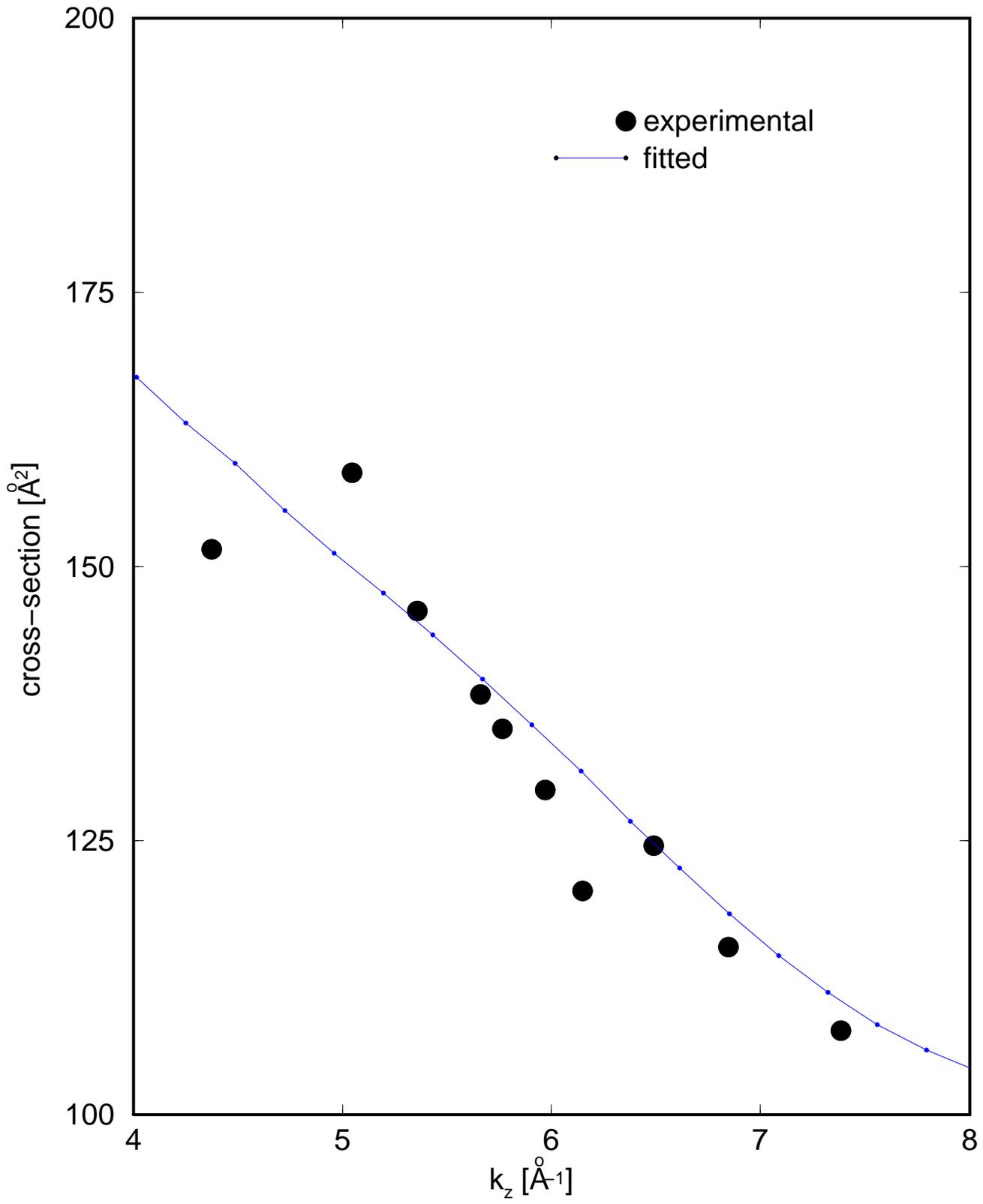
\begin{figure}
\caption{
A comparison between experimental and theoretical cross-sections for He
scattered from an isolated Ag atom adsorbed on a Pt(111) surface. The
experimental values were determined from adsorption curves $I(\Theta)$
(Eq.(\protect\ref{eq:I})) recorded at 38K and extrapolated to $\Theta
\rightarrow 0$. A fit (solid line) of the theoretical cross-sections was used 
to derive the potential used in this work.
}
\label{fig:cs}
\end{figure}

The experiments reported here (e.g., Fig.\ref{fig:cs}) involve two
different modes of operation of the He-scattering apparatus. In the first
mode, the specularly scattered He-intensity is recorded during the deposition
of silver at a constant rate. From the initial slope (extrapolated towards
silver coverage $\Theta \rightarrow 0$) the scattering cross-section of the
Ag-nuclei initially formed on the surface are obtained. From a detailed
analysis of the shape of the deposition curves, we can infer that at the
temperature of 38K at which the data in Fig.\ref{fig:cs} were taken these
initial Ag nuclei are individual Ag atoms randomly distributed on the Pt(111)
surface. The initial slope of the He-intensity curve, therefore, directly
yields the He cross section $\Sigma(k_z)$ (as defined by Eq.(\ref{eq:I})
below), where $k_z = k\,\cos(\theta_i)$ is determined by the wavevector $k$ of
the incident He beam, related to its energy by $E=\hbar^2
k^2/(2m)$. $\theta_i=\theta_f$ (specular scattering) denotes the incident and
outgoing angle of the He beam measured against the surface normal. To measure
the dependence of the cross section $\Sigma$ on $k_z$, either the total
scattering angle $\chi = \theta_i + \theta_f$ or the energy of the incident He
beam has to be varied. We have chosen the second option which, in
practice, we achieve by varying the nozzle temperature from liquid nitrogen
temperature to above room temperature at constant total scattering angle $\chi
= 90^\circ$. This temperature variation corresponds to a change of the energy
$E$ from 18 to 90 meV, i.e., from $k_z = 4.15 \AA^{-1}$ to $9.28 \AA^{-1}$.

The second type of experiment is the analysis of diffraction profiles as
shown, e.g., in Fig.\ref{fig:T-vs-E}. Here, the He-intensity is recorded as a
function of the {\em parallel} wavevector transfer $\Delta{\vec K}$. In the
present case, the incident and outgoing beams lie in the same plane as the
surface normal ({\em in plane scattering}) and $\Delta K = k(\sin\theta_f -
\sin\theta_i)$. The azimuthal orientation of the scattering plane relative to
the surface crystallographic directions can be varied by rotating the crystal
around its surface normal. In Fig.\ref{fig:T-vs-E} the surface was oriented
along the [11\={2}]-direction (i.e., the direction along which the first order
Bragg peaks of the hexagonal substrate lattice are expected).\\ Our scattering
apparatus is equipped with a time-of-flight (TOF) spectrometer allowing to
separate the elastic from the inelastically scattered He intensity by counting
only those He-atoms within a narrow energy window of about 0.5meV centered
around the incident beam energy. In this way, even the very small elastic
signals away from the Bragg peaks can be discriminated. As shown in the next
sections, these elastic features provide important information on the surface
morphology.

All the experiments reported here were conducted at {\em low} temperatures,
$T=38K$. As already mentioned, this ensures that small silver clusters are
formed as a consequence of the limited mobility of the Ag adatoms on the
Pt(111) surface. Although similar in size and distribution, these
low-temperature structures should not be confused with those obtained after
depositing or annealing Ag at {\em high} surface temperature, $T \geq$
620K.\cite{2a,2b,3} At these elevated temperatures, the Ag atoms are embedded
into the topmost Pt(111) layer forming small, stable clusters as a consequence
of the surface strain.\cite{3,8a}. Instead, the low temperature structures
presented here are diffusion limited aggregates composed of Ag atoms adsorbed
{\em on top} of the Pt(111) surface. A detailed account on the difference of
these two configurations will be given in a forthcoming paper.\cite{8b}

\subsection{Theoretical Methods}
In modeling the substrate, we assumed a flat supporting surface. Indeed, the
He/Pt(111) equipotential surface is rather smooth, leading to very little
off-specular scattered He.\cite{9} However, due to the lattice misfit of about
4\% between Ag and Pt,\cite{8b,9a} the assumption of a flat Pt(111) surface is
only a reasonable first approximation. In addition, in all of our calculations
we assume the Ag/Pt(111) system to be rigid. Note that experimentally the
inelastic contribution can be separated by a TOF spectrometer. As for elastic
scattering, results of calculations using a rigid, non-vibrating surface
system should be useful at least for studying the main qualitative
features. Moreover, for diffraction scattering from crystalline surfaces, the
effect of surface vibrations on the scattering intensities can be represented
approximately by a simple Debye-Waller factor.\cite{9b,9c,17} A similar
description should be successful for the angular intensity distributions
obtained in scattering from an adlayer.

\subsubsection{Kinetic Monte Carlo Simulations of Diffusion Limited Aggregates}
\label{KMC}
We outline here the methods used to simulate the compact and fractal Ag
islands (systems (\ref{compact},\ref{fractal});
Figs.\ref{fig:LCC},\ref{fig:fractals}). There is an extensive literature on
the modeling of adsorption, diffusion and aggregation processes on homogeneous
surfaces.\cite{10} To produce realistic arrangements of adatoms on the surface
we adopted the Kinetic Monte Carlo (KMC) method.\cite{11} The Pt(111) surface
was simulated as a hexagonal lattice of 100$\times$100 unit cells with
periodic boundary conditions. As for the lateral interaction between adsorbed
particles, the assumptions made are critical for the shape and size of the
final clusters formed. We assumed that the interactions between two adsorbates
depend on the number of their nearest neighbors, $n_{NN}$. Thus the energy of
a particle in a given configuration is:

\begin{equation}
E(n) = E_0 + n_{NN}\,\epsilon
\label{eq:Ens}
\end{equation}

\noindent where $E_0=5.2$kcal/mole is the activation energy for diffusion at
zero coverage, and $\epsilon=5.0$kcal/mole is the nearest neighbor interaction
energy. No values were available for the $\epsilon$ parameters for Ag on
Pt(111). Therefore, we adopted the values given by Rosenfeld et al.\cite{4}
for Pt adsorbed on Pt(111). These values, while not more than
semiquantitatively valid, should at least allow for the formation of plausible
aggregates.

The diffusion of adatoms was modeled by random walks over nearest neighbor
sites. The hopping rate, i.e., the transition probability per unit time, of a
particle from site $i$, where it has $n_i$ nearest neighbors, to a neighboring
site with $n_f$ nearest neighbors, is taken as

\begin{equation}
\omega_{i \rightarrow f} = \nu \, e^{{-E(n_i)}/(k\, T)} = \omega_0 \,
e^{{-n_i \, \epsilon}/(k\, T)}
\label{eq:w-if}
\end{equation}

\noindent where $\omega_0 = \nu \, \exp[-E_0/(k\, T)]$ is the hopping
frequency of an ``isolated'' particle; $\tau_0=1/ \omega_0$ is the average
time interval between successive moves on the bare (zero coverage)
surface. Since there is an uncertainty in our data regarding the values of the
activation energy $E_0$ and the interaction energy $\epsilon$, it must be
commented that there is a resulting uncertainty in the temperature scale:
Changing $T$ at constant $E(n)$ is equivalent to changing $E(n)$ at constant
$T$. Note further that according to Eq.(\ref{eq:w-if}), the transition rate
depends only on the initial state. A possible dynamical interpretation of this
model, originally suggested by Uebing and Gomer,\cite{12} is that the rate of
the transition from state $i$ to state $f$, is governed by the energy
difference between the initial state and the transition state. The same
is of course true also for the reverse process (with the final state
replacing the initial state), so that detailed
balance is obeyed. We note that a situation where a rate depends on
the gap between any given initial state and the transition state holds
widely for many activated rate processes. Hopping diffusion involves
overcoming a barrier between the initial and final configurations and
therefore the model seems most reasonable here. The KMC
simulations were performed using the time-dependent Monte-Carlo
scheme.\cite{11} In this scheme, instead of randomly choosing particles and
accepting or rejecting moves according to the given transition probabilities
(as in the ``traditional'' MC simulations), one performs a move in any
attempt, and propagates the time accordingly. More explicitly, we first
calculate the average transition rate $\langle r \rangle = 1/\langle \omega_{i
\rightarrow f} \rangle$ out of state $i$, then randomly sample a given $i
\rightarrow f$ move with probability $\omega_{i \rightarrow f}/\sum \omega_{i
\rightarrow f}$ and, finally, perform this move and record the time elapsed as
$\Delta t = 1/\langle r \rangle$. Further details about this procedure can be
found elsewhere.\cite{10,11}

The simulation starts with two particles diffusing on the surface until a
third particle adsorbs. This period is typically 0.05 sec (for simulating an
adsorption rate of 1ML/500sec), corresponding to about $10^6$ MC steps at room
temperature. Then the three particles diffuse, a fourth one adsorbs, and so on
until the desired coverage is obtained. The configurations attained are not at
equilibrium, but reach a steady state after some time duration, when the
adsorption is terminated. The true equilibrium state corresponds most
probably to a segregated phase, with one large compact island formed
by all the adsorbates. Both the compact and the fractal clusters of
Figs.\ref{fig:LCC},\ref{fig:fractals} were generated according to this
procedure, differing only in the respective temperatures of 500K and 200K.

\subsubsection{He Scattering Calculations}
The scattering intensities were calculated using the Sudden Approximation
(SA),\cite{17,13,14,15,16} which has proved very useful in studies of atom
scattering from defects\cite{19} and from substitutionally disordered rare-gas
monolayers.\cite{7a} On the basis of the experience gained with the SA,
including tests against numerically exact calculations for several model
systems,\cite{18} we estimate that at least the main predictions of the SA
calculations should be reliable for the systems studied here.

The SA takes the following form for the systems studied below. Consider a
particular (static) configuration ${\vec r}=\{{\vec r}_1,...,{\vec r}_N\}$ of
a disordered adlayer system consisting of $N$ atoms with ${\vec r}_i$ denoting
the position of atom $i$. The angular intensity distribution for He scattered
from this adlayer involves an average over all configurations pertinent to the
disorder:\cite{14,15}

\begin{equation}
I({\Delta {\vec K}}) = {1 \over A^2} \biggl\langle \left| \int \int e^{i
{\Delta {\vec K}} \cdot {\vec R}} e^{2i\,\eta_{\vec r}({\vec R})} d{\vec R}
\right|^2 \biggl\rangle .
\label{eq:IQ}
\end{equation}

\noindent In this expression ${\Delta {\vec K}} = {\vec K}'- {\vec
K}$ is the wavevector transfer of the He parallel to the surface, where
$({\vec K},k_z)$ is the incident wavevector, and $({\vec K}',-k_z)$ the final
wavevector of the scattered He atom; ${\vec R}= (x,y)$ denotes the coordinates
of the He atom in the surface plane; $A$ is the area of the surface over which
the integration in Eq.(\ref{eq:IQ}) is performed; $\langle ... \rangle$
denotes the average over all configurations of the quantity in parenthesis;
$\eta_{\vec r}({\vec R})$ denotes the phase shift for He scattering for an
adlayer having a configuration ${\vec r}$. The phase shift is given in the WKB
approximation by\cite{13}

\begin{equation}
\eta_{\vec r}({\vec R}) = \int_{\xi({\vec R})}^{\infty} dz\: \left( \left[
k_z^2 - 2m\,V_{\vec r}({\vec R},z)/\hbar^2 \right]^{1/2} - k_z \right) -
k_z\,\xi({\vec R}) ,
\label{eq:eta}
\end{equation}

\noindent where $m$ is the mass of the He atom and $z$ denotes the distance of
the He atom from the surface plane; $V_{\vec r}({\vec R},z)$ is the
interaction potential between the He atom at position $({\vec
R},z)$. $\xi({\vec R})$ in Eq.(\ref{eq:eta}) is the classical turning point
for the He atom when colliding with the surface at the lateral position ${\vec
R}$, i.e., $\xi({\vec R})$ is the $z$ value for which

\begin{equation}
(\hbar k_z)^2 - 2m \, V_{\vec r}({\vec R},z) = 0
\label{eq:turning}
\end{equation}

\subsubsection{The Interaction Potential}
To determine the interaction of He atoms with the Pt(111) surface with a
single adsorbed Ag atom, we calculated the attenuation of the specular peak
with incident He energy. This attenuation is closely related to the
cross-section of the adatom. The relation between the cross-section $\Sigma$
and the specular intensity $I$ for scattering from random {\em isolated}
adatoms is:\cite{19,19b}

\begin{equation}
\Sigma = {1 \over I_0} \lim_{\Theta \rightarrow 0} {dI \over d\Theta} ,
\label{eq:I}
\end{equation}

\noindent where $I_0$ represents the specular scattering intensity from the
clean Pt(111) surface, $n$ is the number of sites per unit cell, and $\Theta$
the adatom coverage. The parameter $\Sigma$ in Eq.(\ref{eq:I}) can be
interpreted as the cross-section of a single adatom. Indeed, Eq.(\ref{eq:I})
is only valid when the cross-section of different atoms can be viewed as
non-overlapping. $\Sigma$ is very sensitive to the interaction
potential. Therefore, we tried to find a potential that is capable of
reproducing the experimentally observed changes in the specular scattering. We
assumed the following general form for the interaction potential between He
and the Pt(111) surface with an adsorbed Ag atom:

\begin{equation}
V_{\rm {He/Ag/Pt(111)}} = V_{\rm{He/Pt(111)}} + V_{\rm{He/Ag}} ,
\label{eq:V_He}
\end{equation}

\noindent where $V_{\rm{He/Pt(111)}}$ is the interaction potential of He with
the clean Pt(111) surface. A He-Pt Morse potential,

\begin{equation}
V(z) = D_e \left( e^{-2\alpha\,(z-z_m)}-2e^{-\alpha\,(z-z_m)} \right)
\label{eq:Vz}
\end{equation}

\noindent for this interaction\cite{19a} is available in the literature, with
$D_e = 7.86$meV$ = 2.89\times10^{-4}$a.u. and $\alpha = 0.98\AA^{-1} =
0.52$a.u.\cite{20} $V_{\rm {He/Ag}}$ is the interaction between He and an
isolated adsorbed Ag atom. We modeled this interaction with a Lennard-Jones
potential, and changed its parameters until the potential could reproduce the
dependence of the experimental cross-section on the incidence energy of the He
atom. This procedure also yielded the value of $z_m=11.46$a.u. in the Morse
potential. The high sensitivity of this fit to the parameter values is the
source of our confidence in the semiquantitative reliability of the potential.

For the interaction between He and multiple adsorbed Ag atoms, we assumed
pairwise interactions for the He atom with the adsorbed Ag atoms, i.e.,
$V_{\rm{He/Ag}}$ in Eq.(\ref{eq:V_He}) is replaced by $\Sigma_i
V_{\rm{He/Ag}}(r_i)$ where $r_i$ is the distance of the He from the i$^{\rm
th}$ Ag atom.  The pairwise additive potential for the interaction with a
collection of Ag atoms is expected in general to be no better than a first
approximation. We estimate, however, that it is sufficient in accuracy for our
purpose here.

\section{Results and Discussion}
\label{results}

\subsection{He Interaction Potential with Pt(111) and adsorbed Ag}
\label{potential}
Specular scattering intensities were measured as a function of He incidence
wavenumber for a {\em single} adsorbed Ag atom. Using these values in
Eq.(\ref{eq:I}) provided us with the experimentally determined cross-sections,
shown in Fig.\ref{fig:cs}. Experimental constraints detailed in
Sec.\ref{experiment} limited the range of incident He wavenumber
values. Subsequently, we attempted to fit the experimental cross-section data
by finding an optimal set of $C_{12}$ and $C_6$ values for the He/(adsorbed Ag
atom) Lennard-Jones 6-12 potential

\begin{equation}
V_{\rm {He/Ag}} = {C_{12} \over r^{12}} - {C_6 \over r^6} ,
\label{eq:LJ}
\end{equation}

\noindent with which the specular intensities were calculated from
Eq.(\ref{eq:I}). By adopting this trial and error approach, we found that the
experimental values are best reproduced with the following Lennard-Jones
parameters: $C_{12}=6.33\times10^6$a.u. and $C_6=29.0$a.u. This leads to
$r_{eq} = (2C_{12}/C_6)^{1/6} = 4.6\AA$ and $V(r_{eq}) \approx 0.9$meV. A comparison between the
experimental and theoretical values is presented in Fig.\ref{fig:cs}. We found
the fit to be very sensitive to these parameters, suggesting that our
empirically determined potential models the long range attractive and the
short range repulsive interactions well, and its accuracy is mainly limited by
the experimental uncertainty.

\subsection{Angular Scattering Patterns from Disordered Overlayers}
The results will be outlined by examining the angular distributions of each of
the disorder classes, and comparing among them. In each case where random
ensembles are discussed, we averaged over 30 configurations.

\subsubsection{Scattering from a Full Monolayer (system
(\protect\ref{monolayer}))}
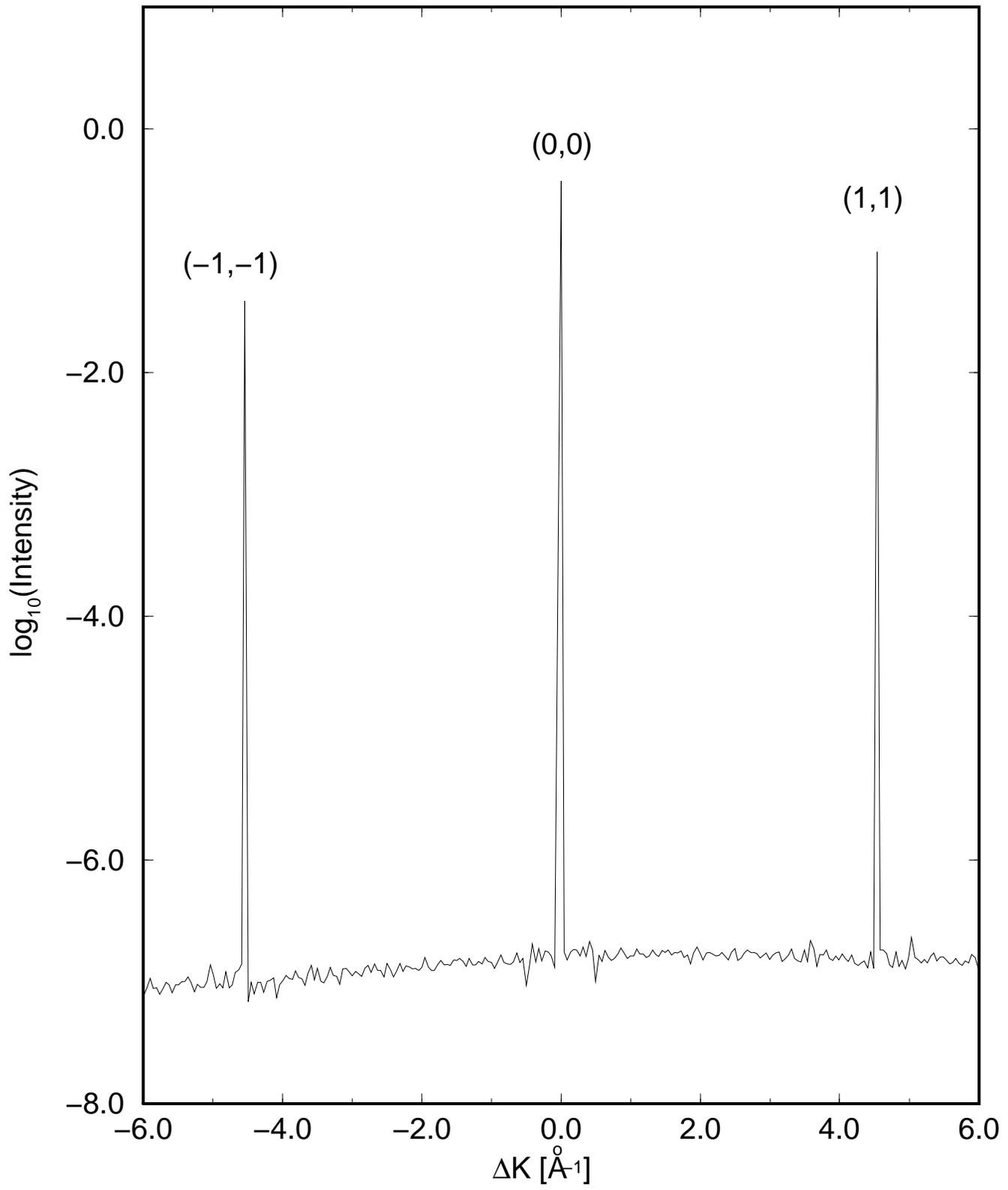
\begin{figure}
\caption{
Calculated angular intensity distribution along the $\Gamma_K$ direction for
He scattered from an ordered monolayer of Ag on the Pt(111) surface, with the
Ag/Pt(111) lattice mismatch ignored. The orientation of the scattering plane
is along the close packed [1\protect\={1}0] direction in real space. The
reciprocal space lattice is rotated by 30$^\circ$ so that the Bragg peaks
appear at non-close packed multiples of $4\pi/a$, $a=2.77\AA$ being the
Pt(111) lattice constant. For the indexing of the Bragg peaks, we chose real
space unit vectors $\protect\vec{a}_1 = (a/2,\protect\sqrt{3}a/2)$,
$\protect\vec{a}_2 = (-a/2, \protect\sqrt{3}a/2)$, corresponding to reciprocal
space unit vectors $\protect\vec{b}_1 = (2\pi/a)(1,1/ \protect\sqrt{3})$,
$\protect\vec{b}_2 = (2\pi/a)(1, -1/ \protect\sqrt{3})$.
}
\label{fig:monolayer}
\end{figure}

To evaluate the influence of surface disorder on He scattering, we first
calculated for reference the angular intensity pattern for He scattered from a
{\em {completely ordered monolayer}} of Ag on Pt(111), i.e., for a silver
coverage $\Theta=1$. (The issue of incommensurability does not arise here
because we take the Pt(111) surface to be flat -- Eq.(\ref{eq:Vz})). The
incident perpendicular He momentum is 3bohr$^{-1} = 5.67\AA^{-1}$ (For the
other calculations presented below, we assumed the same incident momentum.)
The result, presented in Fig.\ref{fig:monolayer}, strong first order Bragg
peaks along with the specular peak. The very weak background intensity is due
to finite grid effects. Note that our calculations, based on a pairwise
additive potential, probably give the effect of a too corrugated surface
(hence diffraction stronger than the real one) since the smoothing role of the
conduction electrons is not included in the interaction.

\subsubsection{Scattering from Single Clusters: Adatom and Small Compact
Cluster (systems (\protect\ref{single}),(\protect\ref{heptamer}))}
\label{single-clusters}

The simplest deviation from perfect order is the introduction of a single
defect. Such defects, namely a single Ag adatom and SCC are the subject of
this section. In subsequent sections gradually more disordered systems will be
considered.

The dashed lines in Figs.\ref{fig:adatom},\ref{fig:heptamer} present the
scattering results from a single adatom and an SCC (heptamer)
respectively. Despite their quantitatively different appearance, the
scattering patterns share some important qualitative features:

\paragraph{Specular Peak:}
The sharp specular peak is due to He scattered from uncovered Pt(111) surface
areas, where the He/Pt(111) potential is not influenced by the defect. The
flatness of this potential across the unperturbed surface results in almost
perfect mirror scattering.
\paragraph{Broadening of the Specular Peak:}
Immediately below the specular ``spike'' a shoulder appears, which is clearly
wider in the single adatom case. In other words, the broadening of the
specular peak is inversely related to the cluster size.
\paragraph{Off-Specular Peaks:}
A series of rather broad, off-specular diffraction features follows the
specular broadening. The interesting feature is their spacing, which is larger
for the single adatom. Again, this spacing in reciprocal space is inversely
related to the cluster size. For a detailed quantitative analysis of this
issue see Ref.[$\!\!$~\onlinecite{7b}]. Here it will suffice to mention that
the peak structure, as well as the specular broadening, can be satisfactorily
explained in terms of scattering from a hard hemispherical object of radius
$d$. The resulting {\em Fraunhofer} diffraction intensities are given
by:\cite{21}

\begin{equation}
I(\phi) \propto \left| {{(1+\cos\phi)\,J_1(k\,d\,\sin\phi)} \over {\sin\phi}}
\right|^2 ,
\label{eq:Fraunhofer}
\end{equation}

\noindent where $\phi$ is the scattering angle, $J_1$ denotes the Bessel
function of first order, and $k$ is the incident wavenumber. The parameter $d$
can be interpreted as the radius of the scattering cross-section. Indeed, we
found using Eq.(\ref{eq:Fraunhofer}), that for scattering from an isolated
heptamer (Fig.\ref{fig:heptamer}), the Fraunhofer interferences are those
characteristic for an average island scattering cross-section of three
nearest-neighbor Ag atoms.

Another mechanism at work in the scattering from defects is {\em rainbow
scattering}, in which the positions of the peaks are determined by the
inflexion points of the equipotential surface.\cite{16,17} For an extensive
treatment of this issue in the heptamer case, see
Ref.[$\!\!$~\onlinecite{7b}].

\subsubsection{Scattering from Adatoms and SCCs Randomly Adsorbed on a Lattice
(systems (\protect\ref{lattice-random}),(\protect\ref{random-heptamers}))}
\label{random-clusters}

A natural transition to more complex disorder is obtained by considering a
collection of clusters, randomly adsorbed on the hexagonal lattice. A system
like this can be formed in realistic conditions when gas phase Ag atoms are
deposited on a cold Pt(111) surface, such that the Ag mobility is very
small.\cite{2a,2b,3,4} Scattering from such {\em translationally random}
systems is the subject of the present section.

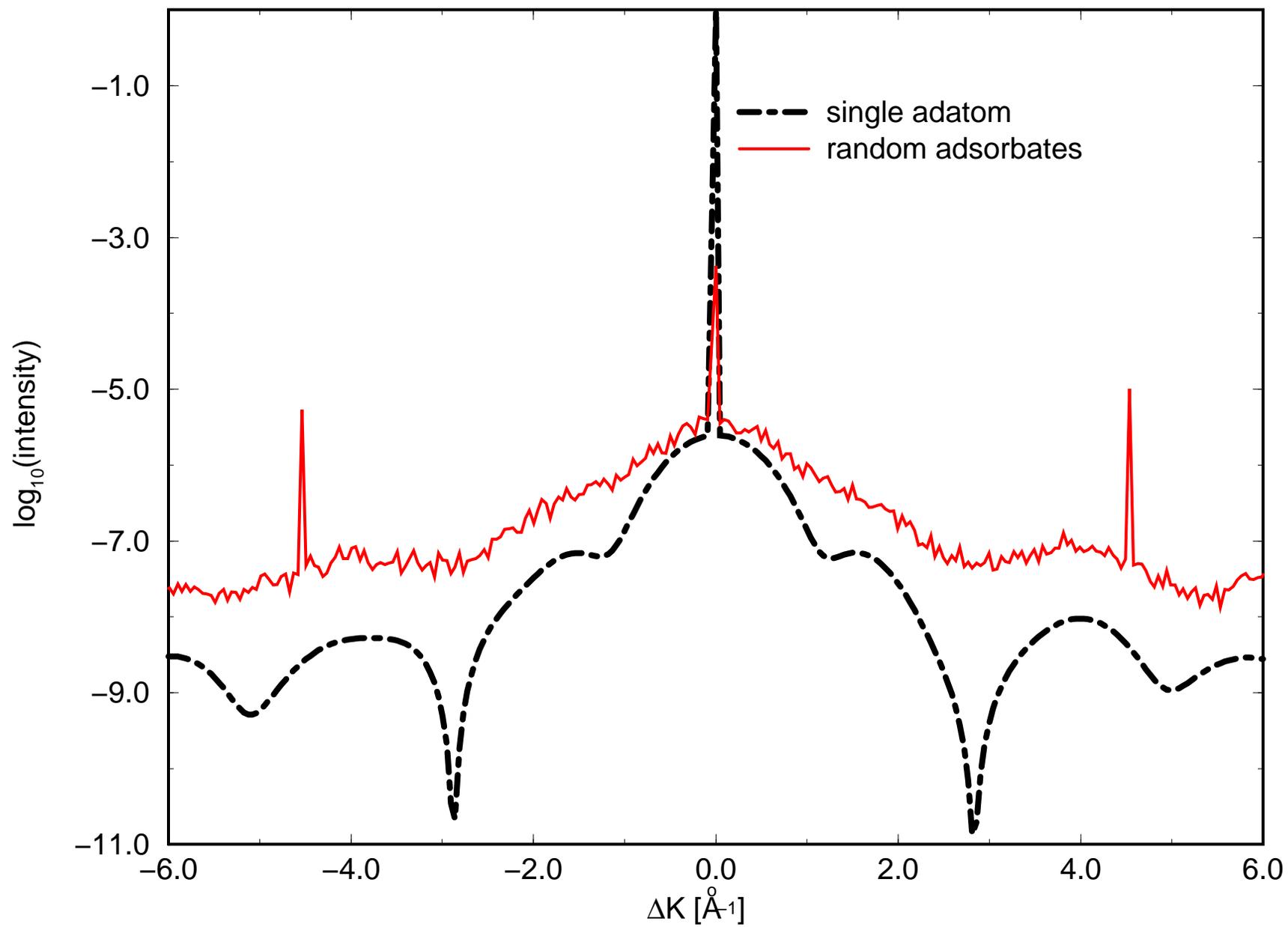
\begin{figure}
\caption{
Dashed line: Calculated angular intensity distribution along the $\Gamma_K$
direction of He scattered from a single Ag adatom. Solid line: Same for He
scattered from 15\% coverage translationally random Ag atoms on a Pt(111)
lattice.
}
\label{fig:adatom}
\end{figure}

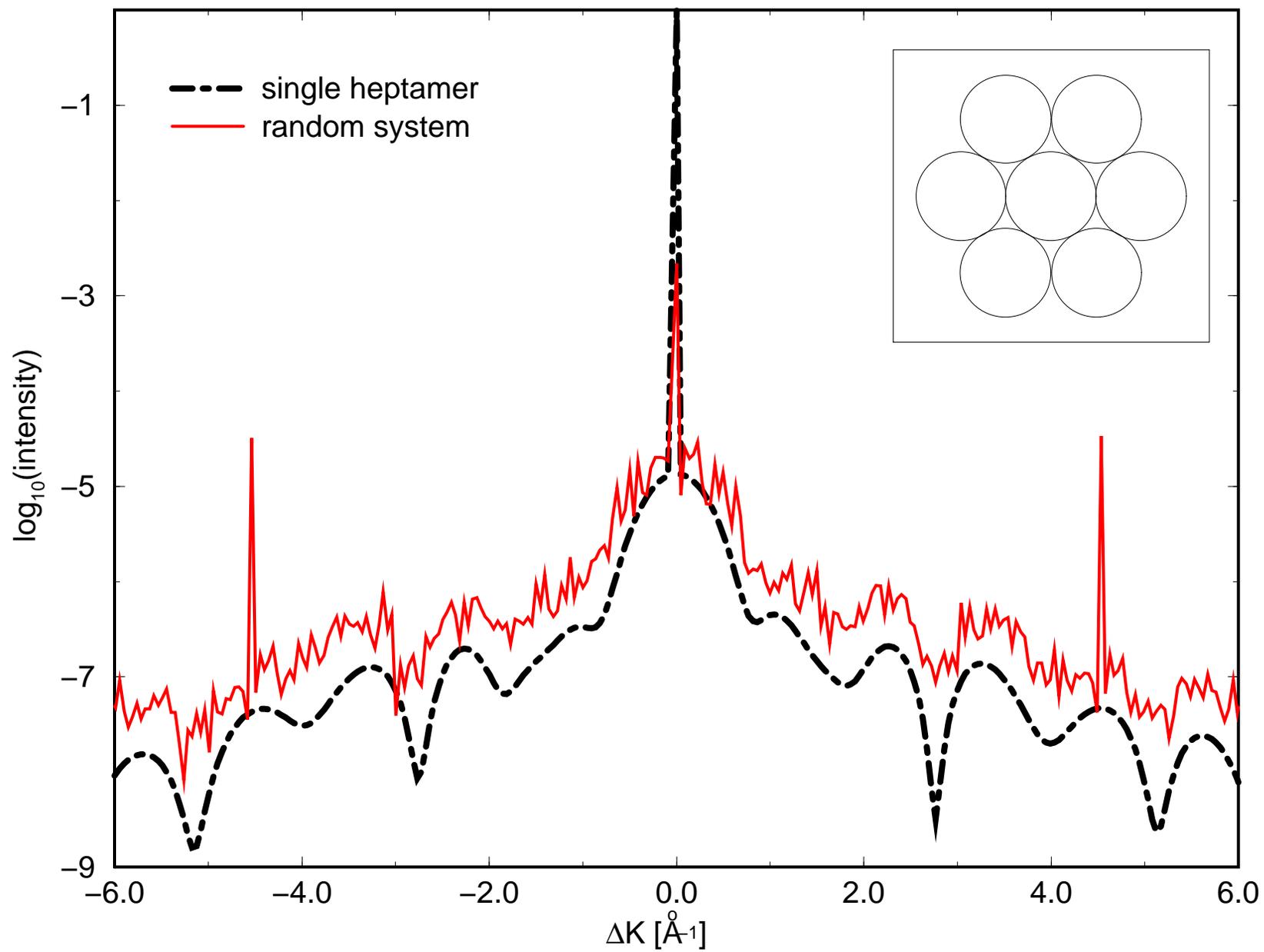
\begin{figure}
\caption{
Dashed line: Calculated angular intensity distribution along the $\Gamma_K$
direction of He scattered from a single Ag heptamer. Solid line: Same for He
scattered from randomly distributed Ag heptamers on Pt(111). Inset: Top view
of a small compact cluster (heptamer).
}
\label{fig:heptamer}
\end{figure}

The results for a system of single adatoms and SCCs at 15\% coverage are
presented as the solid lines in Figs.\ref{fig:adatom},\ref{fig:heptamer}. Let
us enumerate the central features:

\paragraph{Specular and Bragg Peaks:}
The sharp specular peak is as usual an indication that significant portions of
the Pt(111) surface remain flat. In addition one observes strong first and
second order Bragg peaks at multiples of $2\pi/a$ ($a=2.77\AA$ is the Pt(111)
lattice constant). What is their origin? It is not the underlying {\em flat}
Pt(111) surface, as proved by the absence of Bragg peaks in the case of the
individual defect systems of Figs.\ref{fig:adatom},\ref{fig:heptamer}. Nor is
it the small (111) plateau on top of the SCCs (see inset of
Fig.\ref{fig:heptamer}), since the Bragg peaks are observed also for the
adatom system, which has no such structure. Thus these peaks can only be the
result of the strong corrugation induced by the presence of the Ag adatoms and
clusters on the {\em lattice}. This must be so in spite of their {\em random}
positions on this lattice. Since the Bragg peaks are robust and their
intensities are comparatively high ($\sim 10^{-3}$ relative to specular), they
are easily measurable. In addition, they yield to a simple theoretical
description, and thus are attractive candidates for the calibration of a
simple model potential from experimental data.

\paragraph{Similarity to Single Cluster and Effect of Disorder:}
Perhaps the most striking result from the present calculations is the close
(qualitative) resemblance between the intensity distributions from the single
clusters and their random counterparts. This can be seen very clearly in
Figs.\ref{fig:adatom},\ref{fig:heptamer}, where the distributions are slightly
offset for clarity. As is especially conspicuous in the SCC case
(Fig.\ref{fig:heptamer}), there is a one-to-one correspondence between
essentially every (broad) off-specular peak in the single and random case,
respectively. The effect of the randomness seems to be limited to a
modification of the single-cluster intensity distribution by the addition of
noise. In the previous section we identified the principal cause for the peak
structure in the scattering from a single cluster to be Fraunhofer
scattering. This, then, must also be the dominant mechanism in the scattering
from the random systems. The Fraunhofer mechanism is sensitive to {\em local}
surface details, through the cross-section of the individual cluster
[Eq.(\ref{eq:Fraunhofer})]. Since this is a feature of the adlayer which is
not expected to change significantly in the transition to an ensemble, the
similarity agrees with intuition. But why does the randomness have such little
effect? This can be understood qualitatively as follows: For a regular {\em
super-lattice} of islands, one would expect a set of Bragg peaks at multiples
of the inverse super-lattice constant. For a random set of islands, as in the
present case, such long-range translational order is not present, and the only
order that remains in the system is the structure of the individual islands
and the discreteness of the underlying lattice giving rise to the Bragg peaks
at the Pt positions. Hence this is the only coherent contribution (though the
Fraunhofer or rainbow mechanisms) to the intensity spectrum. The effect of the
disorder is then reduced to the addition of noise. In the next section we
investigate what happens when even the pinning due to the underlying substrate
lattice is lost.

\subsubsection{Scattering from a Fully Random Adatom System (system
(\protect\ref{fully-random}))}
\label{sec:fully-random}

In the previous section we considered a random adatom system on a
lattice. Here we consider the same system of randomly located adatoms, with
the lattice constraint removed. Thus, each Ag adatom is located completely
randomly, as if the Pt(111) surface were perfectly smooth. Although this
system is rather artificial, it removes the distance length scale between the
atoms, which the lattice implies. Consequently in this model, adatoms can be
``merged'' into each other into statistically allowed structures. This
non-lattice system is amenable to analytical modeling, as discussed below.

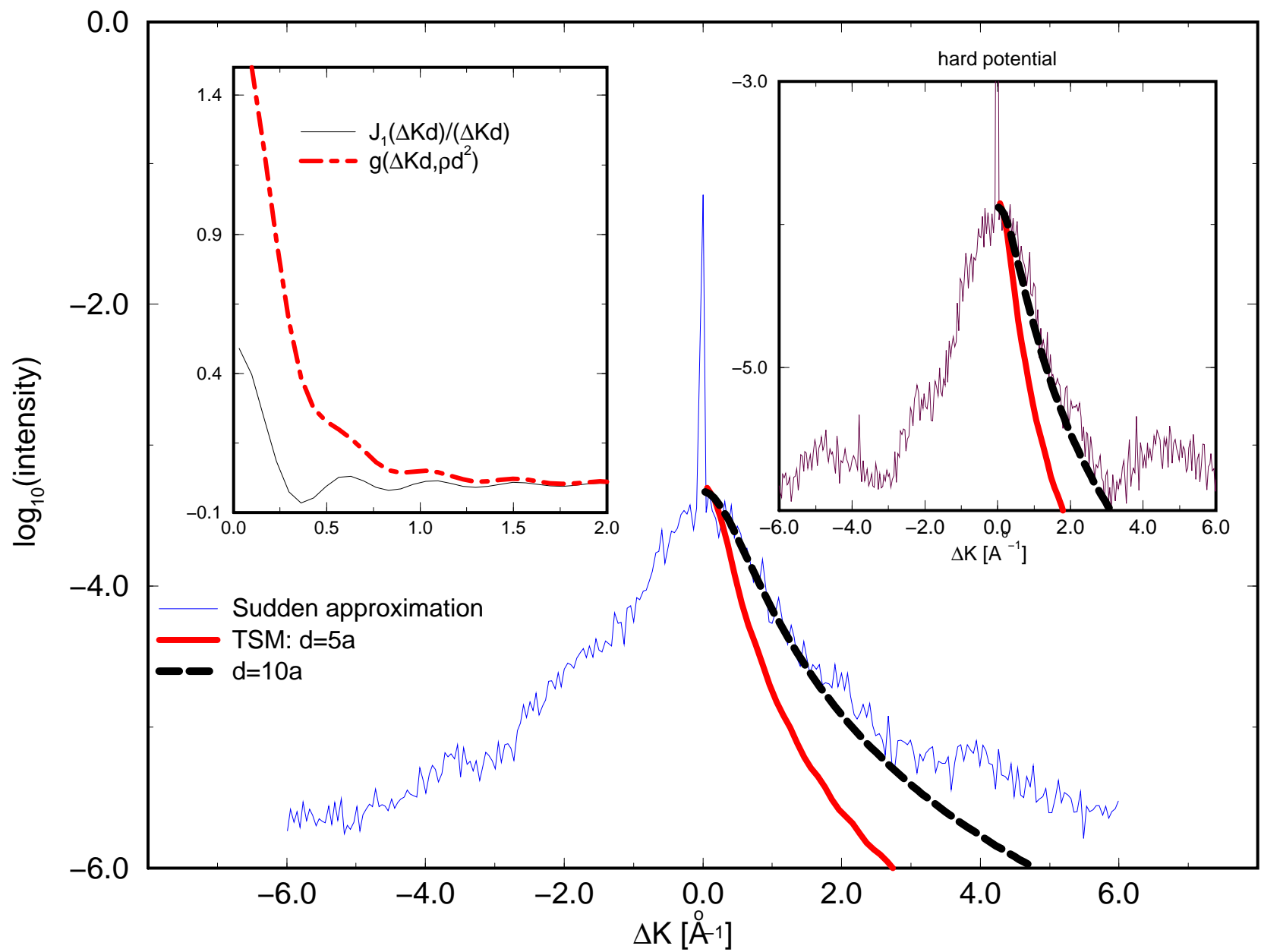
\begin{figure}
\caption{
Calculated angular intensity distribution of He scattered from a Pt(111)
surface with 15\% adsorbed Ag atoms with off-lattice translational
randomness. Thin line: Sudden approximation results. Thick lines: calculation
using Eq.(\protect\ref{eq:P-random-final}). Right Inset: Same, using a harder
He/Ag potential. Left inset: The disorder term $g$ compared to the single
adatom term, from Eq.(\protect\ref{eq:P-random-final}) (linear scale).
}
\label{fig:fully-random}
\end{figure}

The scattering results from this system are presented in
Fig.\ref{fig:fully-random}, for a system of 15\% coverage Ag on Pt(111). The
scattering intensity is very similar to that obtained for the lattice system
of Fig.\ref{fig:adatom}. The broadening of the specular peak is essentially
identical, and the decay of the intensity beyond the broadening is the same
apart from the oscillations absent in Fig.\ref{fig:fully-random}. These
oscillations, then, can most likely be identified as the result of Fraunhofer
interference, whose effect is suppressed due to the loss of adatom identity in
the present case.

Another sharp feature evidenced in Fig.\ref{fig:fully-random} is the presence
of {\em two distinctly different slopes} dominating the intensity
distribution. The slope discontinuity occurs at about $\Delta K^* = 3
\AA^{-1}$, and is an indication that there exist
two important structural regimes on the surface: The individual cluster, and
the long-range translational randomness. Why 3$\AA^{-1}$? The answer can be
found in the single adatom scattering intensity of Fig.\ref{fig:adatom}:
3$\AA^{-1}$ is exactly the point where the specular broadening reaches a
minimum, and the Fraunhofer interference pattern takes over (the actual value
of course depends on the cluster radius). One would expect that the small
$\Delta K$ ($< \Delta K^*$) regime corresponds to the long range
surface-structure, and vice versa. To check this, we repeated the calculation
with a ``harder'' He/Ag potential ($C_{12} = 3.0\times 10^7$a.u., $C_6 =
7.0$a.u.), which enhances the effect due to the individual cluster
structure. The result is presented in the right inset in
Fig.\ref{fig:fully-random}. The slope discontinuity at $\Delta K^*$ is even
more pronounced here, and oscillations reminiscent of the on-lattice system
are observed as well for $\Delta K > \Delta K^*$. The reason is that this
harder potential effectively increases the local surface corrugation;
consequently the surface is rougher and leads to increased high-angle
scattering. Thus changes in the He/Ag potential strongly affect the intensity
distribution for {\em large} $\Delta K$ values, which therefore contain
important information on the He/adsorbate potential. Furthermore, by studying
the large $\Delta K$ regime, a He scattering experiment can be used as a tool
to measure adatom electron densities, which are related to the He/adsorbate
potential.\cite{20a} It is clear from the above analysis that the short-range
structure, manifested by the local corrugation (and controlled by the hardness
of the potential), appears in the {\em large} $\Delta K$ regime.  However, the
small $\Delta K$ behavior is identical even for the hard-potential case. This
proves that this regime is determined exclusively by the long-range structural
features of the surface, as indeed expected.

The question arises whether we can isolate the long-range effect of the
surface disorder ($\Delta K < \Delta K^*$) from that of the interaction with a
single defect.  In order to completely isolate the effect of the translational
randomness of the clusters, we introduce a highly simplified {\em two-state}
model (TSM), in which every adatom is represented by a cylinder of height $h$
and a diameter $d$, and the adatoms are completely randomly located. Here $h$
is the height of an adatom above the Pt(111) surface and $d$ is the diameter
of the adatom scattering cross-section. The adatoms are assumed to be fully
mutually penetrable. The resulting system may be considered as a {\em
two-phase random medium}, in the sense that the underlying surface and the
incomplete layer of tops of islands are two separate phases from the point
of view of the incident He beam. This approximation is reasonable for systems
of metal atoms adsorbed on metal surfaces, because the corrugation of metal
islands is extremely weak. By representing the adatoms as structureless
cylinders, we capture only the {\em global} features of the system, neglecting
its local structure, and can thus hope to reproduce only those features of the
spectrum which result from the underlying random scattering system.

An outline of the solution of this TSM for the intensities is
given next; for additional details see the appendix. For scattering from a hard
island on a hard surface, the general expression for the scattering
intensity [Eq.(\ref{eq:IQ})] becomes:

\begin{equation}
I(\Delta{\vec K}) = {1 \over A^2} \int d{\vec R}_1 \: \int d{\vec R}_2 \: e^{i
\Delta{\vec K} \cdot ({\vec R}_1 - {\vec R}_2)} \langle e^{2i k_z \,Z({\vec
R}_1,{\vec R}_2)} \rangle ,
\label{eq:IQ-new}
\end{equation}

\noindent where

\begin{equation}
Z({\vec R}_1,{\vec R}_2) = \xi({\vec R}_2)-\xi({\vec R}_1)
\label{eq:Z}
\end{equation}

\noindent and $\xi({\vec R})$ is the classical turning point at ${\vec R}$ for
a hard surface, and

\begin{equation}
\langle e^{2i k_z \,Z({\vec R}_1,{\vec R}_2)} \rangle = \int dz\: e^{2i k_z\,z}
f_Z(z;{\vec R}_1,{\vec R}_2) .
\label{eq:average}
\end{equation}

\noindent Here $f_Z(z;{\vec R}_1,{\vec R}_2)$ is the probability density of
observing a height difference $z$ between two points located at ${\vec R}_1$
and ${\vec R}_2$ on the surface. The problem of obtaining $I({\vec Q})$
therefore reduces to the evaluation of $f_Z(z;{\vec R}_1,{\vec R}_2)$. In the
case of realistic shape functions for the island, the evaluation of $f_Z$
is a very hard problem, because it requires to take into account the 3D shapes
of all possible cluster types. The TSM renders the problem of
finding $f_Z$ tractable, at the expense of loosing certain details in the
long-range part of the real intensity distribution.

Rewriting Eq.(\ref{eq:average}) for the case of a stepped surface, we obtain:

\begin{equation}
\langle e^{i\,k_z Z({\vec R}_1,{\vec R}_2)} \rangle = \sum_j e^{2i k_z \,z_j}
{\rm Pr}(z_j;{\vec R}_1,{\vec R}_2)
\label{eq:discrete}
\end{equation}

\noindent where the sum is over all possible height differences, and
Pr$(z_j;{\vec R}_1,{\vec R}_2)$ is the probability to observe a given height
difference between the points ${\vec R}_1$ and ${\vec R}_2$. In the present
case $z_j=0,\pm h$. The problem now reduces to evaluating Pr$(z_j;{\vec
R}_1,{\vec R}_2)$. Our evaluation for randomly adsorbed islands is
presented in the appendix, and shows that for this case the scattering
intensity Eq.(\ref{eq:IQ-new}) becomes:

\begin{eqnarray}
I(\Delta {\vec K}) = c_1(\rho\,d^2,2k_z\,h,A) \delta(\Delta {\vec K}) +
c_2(\rho\,d^2,2k_z\,h,d^2/A) \left[ g(\Delta K d, \rho d^2)
+ {{J_{1}(\Delta K d)} \over {\Delta K d}} \right] \nonumber \\
g(\alpha,\beta) = \int_0^1 J_0(\alpha S) e^{{\frac{1}{2}} \beta \left(
\cos^{-1}S - S \sqrt{1-S^2} \right)} S\,dS ,
\label{eq:P-random-final}
\end{eqnarray}

\noindent where $J_{1}$ is the modified Bessel function of order $1$, and
$\rho$ is the adsorbate density.  The factors $c_1$, $c_2$ are independent of
$\Delta {\vec K}$. The important point to notice about
Eq.(\ref{eq:P-random-final}) is that apart from a specular term, it is
composed of a term due to a {\em single} cylindrical adatom (the $J_{1}$ term
-- compare to Eq.(\ref{eq:Fraunhofer})), and a new term ($g$), which
incorporates the effect of the surface disorder. As can be seen in the left
inset in Fig.\ref{fig:fully-random}, the effect of the disorder term is very
significant: It completely dominates the small $\Delta K$ regime compared to
the single adsorbate term. In particular the {\em slope} in this $\Delta K$
range is controlled almost exclusively by the disorder term (in contrast to
the {\em width}, which was previously shown to be determined by the radius of
the islands). This explains the universality of this slope in all the random
models considered above.

Next, Eq.(\ref{eq:P-random-final}) was applied to study the quantitative
aspects of the small $\Delta K$ region. The height $h$ has mainly the effect
of an overall intensity factor. Since one cannot expect the crude TSM to
correctly predict the intensities, we shifted the intensity curves in the
$(\Delta K,\log(I))$ plane, and focused on the effect of changing the cylinder
diameter $d$. At the incidence wavenumber of $k_z = 3.0$bohr$^{-1}$ =
5.7$\AA^{-1}$, Fig.\ref{fig:cs} predicts a cross-section of $\sim 140\AA^2$,
i.e., a diameter of $\sim 5a$. In Fig.\ref{fig:fully-random} we show the TSM
prediction for $d=5a,10a$. The effect of increasing the diameter is to flatten
the intensity. The $d=5a$ case clearly underestimates the intensity, whereas
for $d=10a$ there is good quantitative agreement in the small $\Delta K$ ($<
\Delta K^*$) regime. Thus, not surprisingly, the TSM does not offer good
quantitative agreement for realistic parameter values. However, the agreement
improves when the TSM is compared to the ``hard-potential'' case (see inset in
Fig.\ref{fig:fully-random}), which is consistent with its formulation as a
hard-wall model.

In conclusion, the relative success of the TSM in representing the small
$\Delta K$ regime (as well as its failure to agree with the large $\Delta K$
region!), demonstrates convincingly that the properties of the long-range
structure of the disordered surface can be found from the quantitative
analysis of this regime.

\subsubsection{Scattering from Large Compact Islands  (system
(\protect\ref{compact}))}
\label{sec:compact}

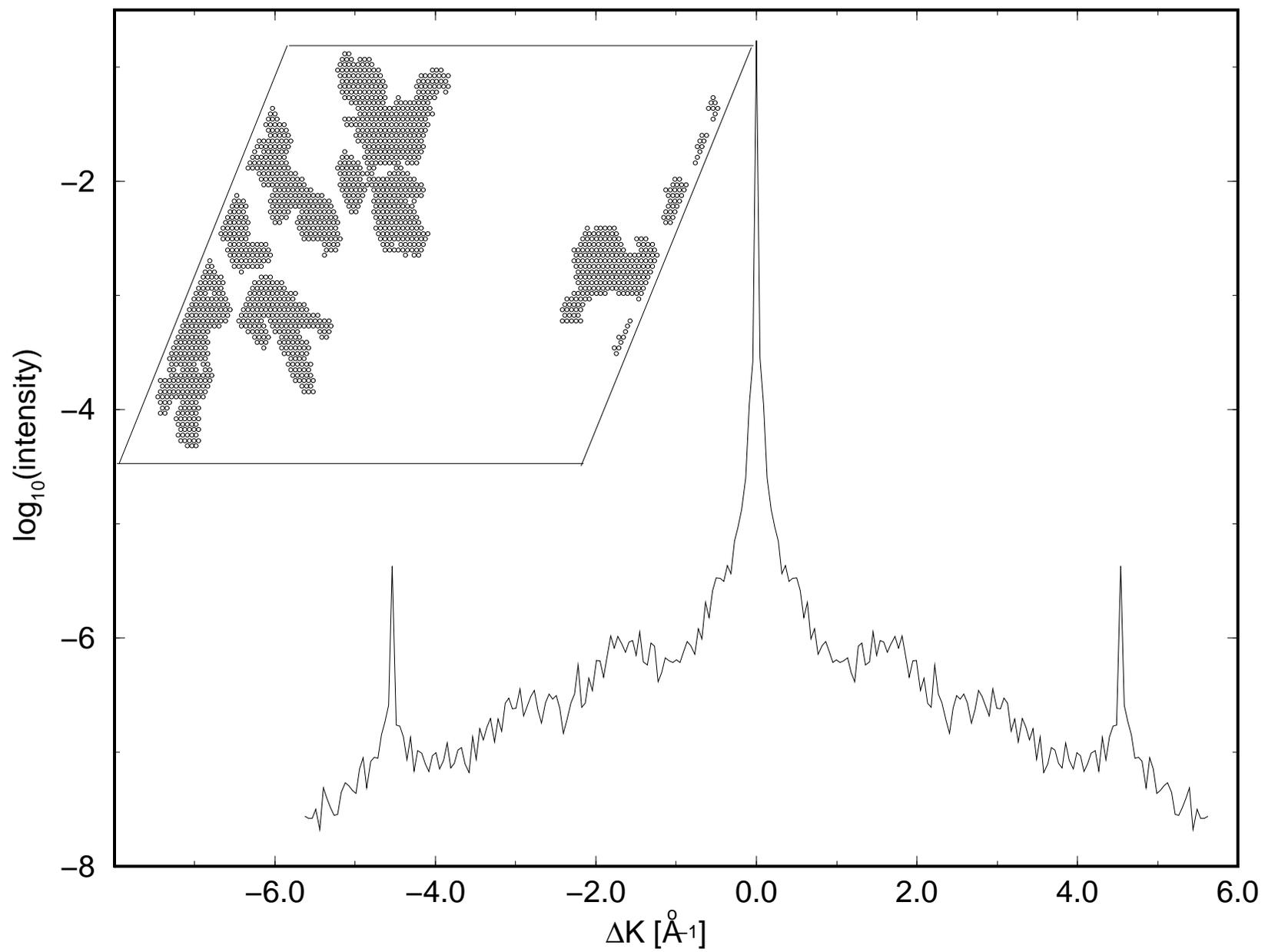
\begin{figure}
\caption{
Calculated angular intensity distribution of He scattered along the $\Gamma_K$
direction from randomly distributed large compact Ag islands on
Pt(111). Inset: Compact Ag islands produced by a KMC simulation, at 15\%
coverage on a surface of 100$\times$100 unit cells and T=500K. The surface is
a parallelogram with an angle of $60^\circ$. Periodic boundary conditions are
employed, so straight edges usually imply that the cluster is continued at the
opposite boundary.
}
\label{fig:LCC}
\end{figure}

In this and the next section we discuss the scattering from yet more complex,
stochastic systems: those created by the Kinetic Monte Carlo algorithm
described in Sec.\ref{KMC}. A typical configuration of the large,
quasi-compact islands formed at 500K is shown in the inset in
Fig.\ref{fig:LCC}. Note that periodic boundary conditions are employed. It is
not our purpose here to obtain a quantitative description of the structure of
these islands from the scattering intensities. Instead, we will limit
ourselves to a discussion of some of the main features of the intensity
distribution, and contrasting them with those characteristic of the systems
studied above.

\paragraph{Broadening of the Specular Peak:}
Among all the systems discussed so far, the specular broadening observed in
Fig.\ref{fig:LCC} is the smallest. This corresponds, as expected, to the large
area of the individual islands. The observed width is $\sim 0.3\AA^{-1}$, in
agreement with an average island diameter of 13-17$a$ obtained in our KMC
simulations. Thus, as indeed obvious, the specular broadening can be used
experimentally to find the average diameter of randomly adsorbed islands.

\paragraph{Broadening of the Bragg Peaks:}
In contrast to the case of translationally random adatoms or SCCs, there is
little off-specular structure in Fig.\ref{fig:LCC}. The exception are the
Bragg peaks, whose existence in the present case is not surprising,
considering the large island surface area, which has the underlying Pt(111)
lattice corrugation. However, the new feature is the {\em broadening} of the
Bragg peaks. The broadening is roughly half that of the specular peak. This
phenomenon is reminiscent of Fraunhofer interference of light transmitted
through a grating. Indeed, roughly speaking, the uncovered surface area
between the islands can be thought of as a slit pattern. The kinetics
governing the formation of the islands leads to depletion zones, or an {\em
effective} repulsion between them (since they cannot overlap), thus inducing a
degree of regularity in this pattern, which can yield the Fraunhofer
broadening. Note that this is different from the Fraunhofer scattering by a
hemispherical object described by Eq.(\ref{eq:Fraunhofer}), which is 
responsible for the specular peak broadening and the ``humps'' at $\Delta K =
\pm 1.5, \, 3.0 \AA^{-1}$. The irregularity
in ``slit'' sizes and orientations, as well as to a lesser extent their
positions, both due to the inhomogeneity of island sizes and separation, is
responsible for the added noise.

\subsubsection{Scattering from Fractal Islands  (system
(\protect\ref{fractal}))}
\label{sec:fractals}

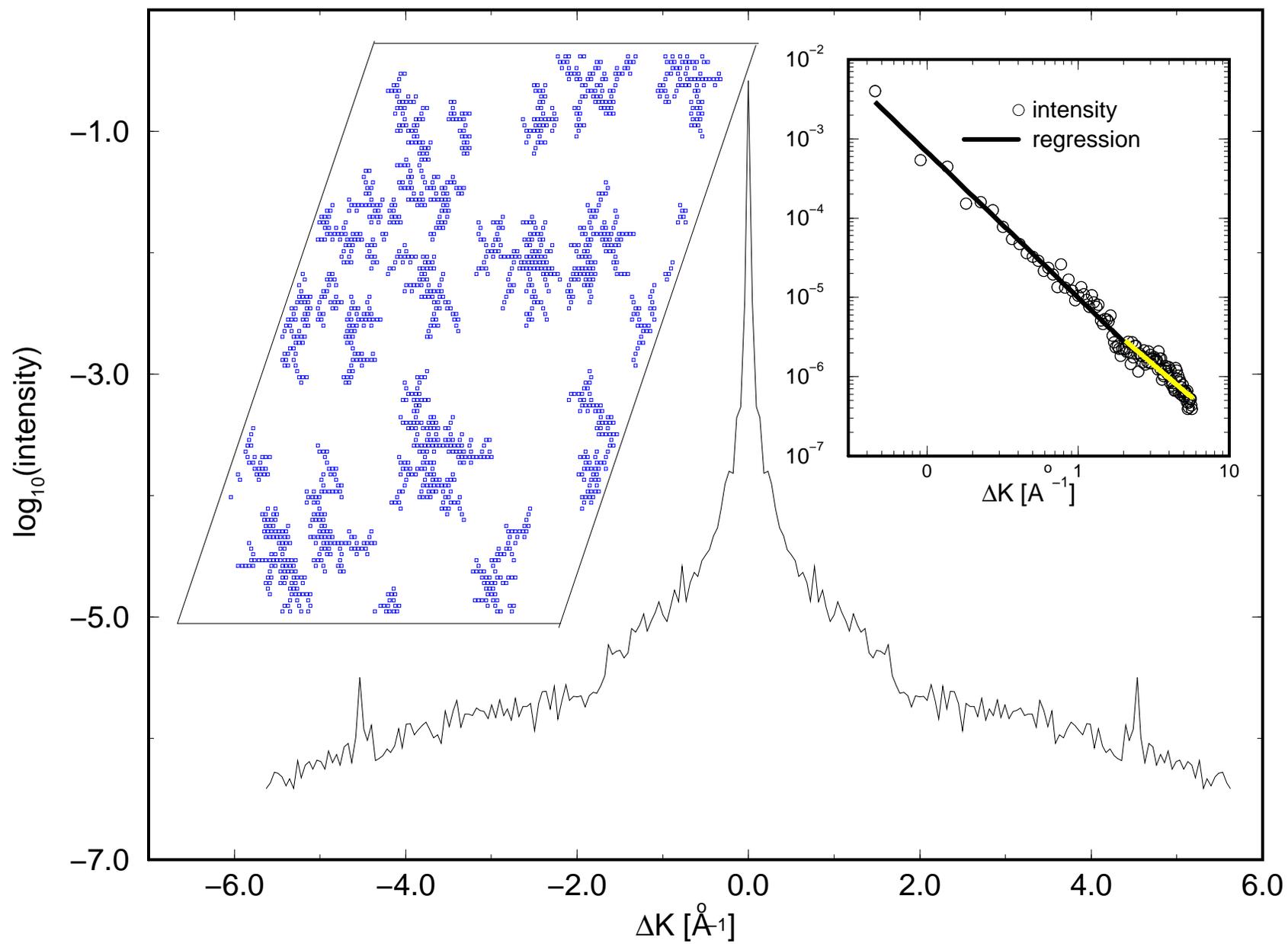
\begin{figure}
\caption{
Calculated angular intensity distribution of He scattered along the $\Gamma_K$
direction from a Pt(111) surface with 15\% adsorbed Ag atoms, forming fractal
islands. Left inset: KMC simulations as in Fig.\protect\ref{fig:LCC}, but for
fractal islands (T=200K). Right Inset: Power-law regression of the intensity
(see text for details).
}
\label{fig:fractals}
\end{figure}

In this section we discuss the scattering results from the last model system
considered in this work: the dendritic/fractal structures formed by the KMC
simulations at 200K. The interest in such structures hardly needs
recapitulation. The averaged scattering intensities, as well as a typical
configuration, are shown in Fig.\ref{fig:fractals}. This configuration is
rather reminiscent of STM results on Pt/Pt(111)\cite{7c,21aaa} and
Ag/Pt(111)\cite{21aa} obtained under similar conditions. The fact that these
systems are indeed fractal over a number of orders of magnitude of
resolution-observation, is discussed in detail elsewhere.\cite{21a}

Interestingly, in spite of the non-negligible surface area of the clusters,
there is apart from weak first order Bragg peaks almost no trace of
interference in Fig.\ref{fig:fractals}. This is in agreement with the
characterization of a fractal as an object with no typical length scale
between physical lower and upper cutoffs. Indeed, our ``fractals'' are
composed of a large variety of clusters of different sizes (dimers, trimers
and larger clusters), which generally are somehow connected. Each of these
clusters contributes Fraunhofer peaks at its typical length scale. The net
effect is to smear out these peaks into a relatively smooth decline of the
intensity.  However, since the fractal clusters are all finite in size, their
average radius {\em is} reflected in the scattering distribution, as the by
now familiar specular broadening. Apart from this, the absence of a typical
length scale (above the lower cutoff $a$) results in that there is essentially
no structure left in the intensity distribution.

As for the intensities at $\Delta K$ beyond the broadening, we observe that
the decline of the angular intensity with $\Delta K$ for fractal islands is
very similar to that for randomly adsorbed adatoms (Fig.\ref{fig:adatom}), and
quite similar to the average decline for SCCs and large compact islands
(Figs.\ref{fig:heptamer},\ref{fig:LCC}). Furthermore, as mentioned earlier,
this dependence is determined by the He/Ag/Pt(111) interaction
potential. Nevertheless, we attempted to fit the intensities both before and
beyond the broadening ($|\Delta K| > 2\AA^{-1}$) by a potential-independent
power-law, following Ref.[$\!\!$~\onlinecite{8}]:

\begin{equation}
I(\Delta K) \propto \Delta K^{D-4} .
\label{eq:power}
\end{equation}

\noindent Here $D$ is the fractal dimension. The right inset of
Fig.\ref{fig:fractals} shows the result of this fit for both positive and
negative $\Delta K$ (the latter were reflected), after elimination of the
specular and Bragg peaks. The result is a slope of $-1.84\pm 0.03$ for
$|\Delta K| < 2\AA^{-1}$, and $-1.62 \pm 0.06$ for $|\Delta K| > 2\AA^{-1}$
over a total range of slightly less than one decade, with respective
regression coefficients of -0.99 and -0.90. If these slopes were used in
accordance with Eq.(\ref{eq:power}), the fractal dimension would be
$D>2$. Clearly, this cannot be for our planar sets, so that we conclude that
the procedure suggested by Eq.(\ref{eq:power}) is {\em not} generally valid
for the determination of fractal dimensions from He scattering data. A further
reason is that the intensities beyond the specular broadening depend strongly
on the hardness of the potential. For example, on changing the He/Ag/Pt(111)
potential, a very different dependence, similar to that in the inset in
Fig.\ref{fig:fully-random}, is obtained (not shown here). Consequently, one
would be forced to conclude that the fractal dimension changes when the
interaction potential is varied. Within a simplistic interpretation, in which
$D$ is the {\em mass} fractal dimension, this is physically unacceptable. If,
however, He measures the fractal dimension of the {\em electron density
contours} -- see Ref.[$\!\!$~\onlinecite{21a}] -- a dependence on the
potential is in fact expected.  For mass fractals, analytical results have
been obtained showing that the scattering intensities are self-affine
functions, whose scaling exponent is simply related to the dimension of the
scattering fractal sets.\cite{22}

\subsection{Comparison of Experiments on Ag/Pt(111) with Scattering From
Theoretical Disorder Models}
\label{experiment-theory}

In this final section we apply the understanding developed from the analysis
of the various disorder models, to interpret He scattering data obtained in
the experiments described in Sec.\ref{experiment}. To recapture the main
details, the system studied in these experiments was a submonolayer of Ag at
50\% coverage deposited on a Pt(111) surface at 38K. He scattering profiles
were taken along the [11\={2}]-direction with incident wavenumber of
6.43$\AA^{-1}$. The experimental data are the circles in
Figs.\ref{fig:T-vs-E}a-c (the same data is shown in all figures). Our purpose
in performing the analysis to be described next was threefold: To {\em
qualitatively} interpret the available experimental data and thus gain some
understanding of the morphology of this low-temperature Ag/Pt(111) phase, to
test the insight developed from the analysis of the disorder models, and to
self-consistently test our empirically extracted potential.

If one focuses attention on the experimental data in, say,
Fig.\ref{fig:T-vs-E}b, several features (some of which observed in the
previous sections) stand out:
\begin{itemize}
\begin{enumerate}
\item{Sharp specular and Bragg peaks, the latter having ``shoulders''.}
\label{spec+Bragg}
\item{Off-specular structure manifested in oscillations of the intensity.}
\label{off-spec}
\item{The {\em absence} (or at least masking) of specular peak broadening;
instead clear maxima at $-0.28, 0.37 \AA^{-1}$ are present.}
\label{maxima}
\item{An asymmetry between positive and negative $\Delta K$.}
\label{asymmetry}
\end{enumerate}
\end{itemize}

The last point is an immediate consequence of the experimental scattering
geometry, which breaks the left-right symmetry (see
Sec.\ref{experiment}). This feature cannot be reproduced in the Sudden
approximation calculations, in which the He always strikes the surface along
its normal.  The other observations can be used to reach several conclusions
on the underlying surface structure:

\noindent (\ref{spec+Bragg}) The large compact clusters model [system
(\ref{compact})] cannot be ruled out since the broadening of the Bragg peaks
is one of its main features.

\noindent (\ref{off-spec}) The presence of a series of peaks does not agree
well with a fractal structure, which should exhibit a rather smooth decay of
the intensity. Instead, one observes in Fig.\ref{fig:T-vs-E}b that the
scattering pattern from a 50\% coverage fractal does contain peaks, for
$\Delta K < 3\AA^{-1}$. The reason is that at such high coverage part of the
fractal character is lost: adatoms fill in the gaps typical of the low
coverage fractals of Fig.\ref{fig:fractals}, and the resulting surface
structure is at first sight not very different from the random one of
Fig.\ref{fig:T-vs-E}a. However, the scattering intensities are markedly
different in Figs.\ref{fig:T-vs-E}a and b: the intensity decay in the fractal
case is far smoother that in the random one, in agreement with the absence of
a typical length scale ($>a$) in the fractal. Point \ref{off-spec}. then, can
be used to eliminate both the fractal and randomly adsorbed adatom models: the
former due to a lack of peak structure for $\Delta K > 3\AA^{-1}$, in the
latter due to a complete disagreement in peak positions.

\noindent (\ref{maxima}) The maxima adjacent to the specular are a feature
which was absent in the disorder models we studied. They are most likely a
result of constructive interference due to a length scale of
$\sim(2\pi/0.33)$, i.e. about 4 lattice constants ($19/(2.77 \sqrt{3})$) along
the [11\={2}] direction. At present we do not yet have a satisfactory model
which contains such a length scale and is also in agreement with the other
features of the intensity distribution.

Having essentially ruled out the fractal and random adatoms models, what then
is an appropriate model which can at least qualitatively explain the features
observed in the experimental intensities?\cite{comment} The central clue in
answering this question is the off-specular peak structure. In the absence of
agreement with the above models, we attempted a fit with a hybrid of the large
compact islands and SCC model, namely a narrow {\em distribution} of
translationally random small compact clusters. The advantage of this model is
that it turns out that the positions and intensities of the off-specular peaks
are extremely sensitive to the relative numbers and absolute diameters of
these clusters. Thus a possible, although definitely not final morphological
identification, is given in Fig.\ref{fig:T-vs-E}c. The distribution shown is a
set of hexagonal islands, 7 or 9 Ag atoms in diameter, mixed in a 1:4 ratio at
50\% coverage. Upon a careful examination it will be noticed that apart from
the quantitative location of the peaks near the specular discussed above, and
the elevated intensity around the first order Bragg, the positions of all
other peaks are correctly reproduced (some of these peaks are mere shoulders
in the experimental data; e.g. at 3.0 and 4.6$\AA^{-1}$). This agreement is
not present in either the random adatom or fractal cases. On the other hand,
the two missing features in the intensity pattern of Fig.\ref{fig:T-vs-E}c
seem likely to be explained by the introduction of some degree of {\em
positional order} in the locations of the SCCs: A separation of about
20-30$\AA$ between the centers of every pair of adjacent clusters would result
in the observed peaks, while at the same time create a shoulder underneath the
Bragg peaks due to a slit-Fraunhofer effect [as for system
(\ref{compact})]. Such an ordered separation is in agreement with the fact
that due to the size-mismatch between Pt and Ag, a strain is induced by the
formation of an Ag island, resulting in an effective repulsion between
neighboring islands. We observe evidence for this in the location of the
experimental first order Bragg peak, which is located slightly below the
position of the corresponding pure Pt(111) peak, suggesting some degree of
lattice ordering due to Ag (which has a lattice constant of 2.89\AA). In
addition there are some purely kinetic effects determining the size and
distribution of islands, namely nucleation and an average diffusion
length. The generation of a size-dispersed SCCs distribution satisfying the
separation requirement turns out, however, to be a non-trivial task, and for
the present qualitative purposes, we limit ourselves to the translationally
random model. Beyond the introduction of positional order, one should probably
also relax the assumption of perfect hexagons: At the low surface temperature
considered, the mobility of Ag adatoms is extremely small, and consequently a
more realistic model is probably one in which the hexagons are imperfect at
least along their perimeter. Since the hexagons are perfect in our model, one
observes an oscillation in the calculated intensity of Fig.\ref{fig:T-vs-E}c,
with a period of $\sim 0.25\AA^{-1}$; this is likely due to interference
between adatoms separated by 4-5 lattice constants along the chains making up
the edges of the hexagons. At any rate, while clearly not being fully
quantitative, the fit with the narrowly size-distributed SCCs is very
promising and bound to be non-accidental.

\section{Concluding Remarks}
\label{conclusions}

The motivation for the study presented in this paper is the rapidly growing
interest in structures formed on surfaces during epitaxial growth. We
demonstrated that various types of disordered structures thus formed can be
studied quantitatively by He scattering, and that a distinction between them
can be made on the basis of the different scattering features that arise
respectively in these cases. This is due to the high sensitivity of the He
angular intensity spectrum to the surface morphology and electron
distribution. In particular, we inverted a He/Ag/Pt(111) potential from
experimental cross section measurements, and subsequently used this potential
to predict from the analysis of experimental angular intensity data the
presence of a phase of small and compact, narrowly size-dispersed clusters
with some degree of positional order. The sensitivity to changes in both
potential and morphology parameters of our fit to the experimental data, as
well as support from a more general He-scattering study\cite{23} and
independent STM data,\cite{Roder} provides strong evidence for the existence
of this phase. Furthermore, it illustrates what we believe to be an important
principle in surface science: The {\em combination} of the mutually
complementary experimental (He scattering and STM) and theoretical techniques
provides an extremely powerful tool for the quantitative analysis of complex
surface structures.

Our theoretical findings shed light on several other points:
\begin{itemize}
\begin{enumerate}
\item{
There are clear and specific differences between the intensity distributions
resulting from the scattering from different classes of disordered
structures. Careful analysis of the relevant features, such as broadening of
the specular peak, the slope ($d\log I/d\Delta K$), off-specular Fraunhofer
maxima, position and width of Bragg peaks and more, can lead a long way to the
identification of the type of disordered phase present on the surface.}
\item{
The structure of the complex intensity spectrum obtained by scattering from
translationally disordered, small compact clusters, can be understood as a sum
of the contributions due the scattering spectrum of a single such cluster, and
due to the disorder. The former is mainly responsible for characteristic
Fraunhofer and rainbow peaks in the large $\Delta K$ regime, whereas the
latter affects the slope of the intensity function for small $\Delta K$ values
and is responsible for the introduction of noise and peak
``smearing''. Conversely, this information can be used to extract useful
information about the individual cluster size and the cluster ensemble
statistics.}
\item{
The claim that for fractals the angular intensity dependence on parallel
momentum transfer follows a universal power law, is not supported by our
calculations. Instead there is a strong dependence on the parameters of the
He/adsorbate potential. Thus it does not appear to be possible to calculate
the fractal dimension from a simple scaling analysis of the intensity
distribution.}
\end{enumerate}
\end{itemize}

Recent experimental studies have addressed the different kinds of disordered
structures formed during 3D epitaxial growth. Specifically, the attenuation of
the He scattering specular peak during growth was measured, allowing for the
distinction between 2D vs. 3D growth mechanisms. The results presented here
indicate that much more information on the structure of disordered multiple
ad-layers can be obtained by He scattering, and we are currently pursuing such
work.

\section*{Acknowledgements}
This work was supported by Grant No. I-215-006.5/91 from the German-Israel
Foundation for Scientific Research (G.I.F.) to R.B.G. and G.C. The Fritz Haber
Research Center at the Hebrew University is supported by the Minerva
Gesellschaft f\"{u}r die Forschung mbH, M\"{u}nchen, Germany. The research
was supported in part by the Institute of Surface and Interface Science at
U.C. Irvine.

\newpage

\appendix
{\bf APPENDIX}

We derive Eq.(\ref{eq:P-random-final}). This requires first the evaluation of
Eq.(\ref{eq:discrete}). The task at hand is the calculation of the
probabilities ${\rm Pr}(z_j; {\vec R}_1,{\vec R}_2)$, to observe a given
height difference between points located at ${\vec R}_1$ and ${\vec R}_2$ on
the surface. Denoting the underlying surface as $0$ and the atom tops as $1$,
there exist four possibilities:\\
(1) ${\vec R}_1,{\vec R}_2 \in 0$ with $Z({\vec R}_1,{\vec R}_2)=0$;\\
(2) ${\vec R}_1\in 0$, ${\vec R}_2\in 1$ with $Z({\vec R}_1,{\vec R}_2)=h$;\\
(3) ${\vec R}_1\in 1$, ${\vec R}_2\in 0$ with $Z({\vec R}_1,{\vec R}_2)=-h$
and\\
(4) ${\vec R}_1,{\vec R}_2 \in 1$ with $Z({\vec R}_1,{\vec R}_2)=0$.\\
Denote $S_{\epsilon_{{\vec R}_1} \epsilon_{{\vec R}_2}}$ (with
$\epsilon_{{\vec R}_i} = 0,1$ according as to whether ${\vec R}_i \in 0$ or
${\vec R}_i \in 1$), as the probability of observing the point ${\vec R}_1$ in
phase $\epsilon_{{\vec R}_1}$ and the point ${\vec R}_2$ in phase
$\epsilon_{{\vec R}_2}$. Then clearly:

\begin{equation}
{\rm Pr}(0; {\vec R}_1,{\vec R}_2) = S_{00} + S_{11} \:\:\:\:\:\:\:  {\rm
Pr}(h; {\vec R}_1,{\vec R}_2) = S_{01} \:\:\:\:\:\:\:\: {\rm Pr}(-h;{\vec
R}_1,{\vec R}_2) = S_{10}
\label{eq:relations}
\end{equation}

\noindent A general formalism for the evaluation of quantities of the type
$S_{\epsilon_{{\vec R}_1} \epsilon_{{\vec R}_2} \cdots \epsilon_{{\vec R}_n}}$
is given by Torquato and Stell (TS).\cite{25} Their result for the case of
interest to us is:

\begin{equation}
S_n({\vec R}_1,..,{\vec R}_n) = 1 + \sum_{k=1}^N {(-\rho)^k \over k!} \int
\cdots \int g_k\left({\vec x}^k\right) \prod_{j=1}^k \left( \left[ 1 -
\prod_{i=1}^n (1-m(|{\vec R}_i-{\vec x}_j|)) \right] \right) d{\vec x}_j
\label{eq:Sn}
\end{equation}

\begin{equation}
S_{11}({\vec R}_1,{\vec R}_2) = 1 - (S_1({\vec R}_1) + S_1({\vec R}_2)) +
S_{00}({\vec R}_1,{\vec R}_2)
\label{eq:S11}
\end{equation}

\begin{equation}
S_{01}({\vec R}_1,{\vec R}_2) = S_1({\vec R}_1) - S_{00}({\vec R}_1,{\vec R}_2)
\label{eq:S01}
\end{equation}

\begin{equation}
S_{10}({\vec R}_1,{\vec R}_2) = S_1({\vec R}_2) - S_{00}({\vec R}_1,{\vec R}_2)
\label{eq:S10}
\end{equation}

\noindent where $S_n$ is shorthand for $n$ points in phase $0$ (so e.g. $S_1$
above means one point in phase $0$), $\rho = N/A$ is the adsorbate density,
${\vec x}^k \equiv ({\vec x}_1,...,{\vec x}_k)$, and $g_k$ is the reduced
distribution function of $k$ particles, defined as:

\begin{eqnarray}
g_k\left({\vec x}^k\right) = {\rho_k\left({\vec x}^k\right) \over \rho^k}\\
\rho_k\left({\vec x}^k\right) = {N! \over (N-k)!}\, P_k\left({\vec
x}^k\right)\\
P_k\left({\vec x}^k\right) = \int \cdots \int P({\vec x}_1,...,{\vec x}_N) \,
d{\vec x}_{k+1} \cdots d{\vec x}_N .
\label{eq:g-def}
\end{eqnarray}

\noindent In the case of an isotropic system $g_2\left({\vec x}_1,{\vec
x}_2\right) = g(r)$ (where $r=|{\vec x}_1-{\vec x}_2|$) is the well-known
radial distribution function. The function $m(|{\vec R}_i-{\vec x}_j|)$ is the
``particle indicator function'', defined as:

\begin{equation}
m(|{\vec R}-{\vec x}_j|) = \left\{ \begin{array}{ll}
	1 \:\:          & \mbox{if $|{\vec R}-{\vec x}_j| < d/2 $} \\
	0 \:\:          & \mbox{if $|{\vec R}-{\vec x}_j| > d/2 $} . \\
	\end{array}
\right.
\label{eq:m}
\end{equation}

\noindent Collecting
Eqs.(\ref{eq:IQ-new}),(\ref{eq:discrete}),(\ref{eq:relations}), we obtain:

\begin{eqnarray}
I(\Delta {\vec K}) = {1 \over A^2} \int d{\vec R}_1\,d{\vec R}_2 \: e^{i
\Delta {\vec K} \cdot({\vec R}_1-{\vec R}_2)} \times \nonumber \\
\left[ (S_{00}({\vec R}_1,{\vec R}_2) + S_{11}({\vec R}_1,{\vec R}_2)) +
e^{2i\,h\,k_z} S_{01}({\vec R}_1,{\vec R}_2) +  e^{-2i\,h\,k_z} S_{10}({\vec
R}_1,{\vec R}_2) \right] .
\label{eq:P-S's}
\end{eqnarray}

\noindent The last equation, together with Eqs.(\ref{eq:Sn})-(\ref{eq:g-def}),
establish a formal connection between the reduced distribution functions of
interest in characterization of surface statistics and the measurable
scattering intensity.

Eq.(\ref{eq:P-S's}) can next be applied to the scattering from translationally
random adatoms. It is shown by TS that
in this case, since $g_k \equiv 1$ for all $k$, one obtains:

\begin{equation}
S_n({\vec R}_1,..,{\vec R}_n) = e^{-\rho\,V_n({\vec R}^n; d)} ,
\label{eq:Sn-random}
\end{equation}

\noindent where $V_n$ is the union area of $n$ circles with diameter $d$, with
centers at $\{{\vec R}_1,..,{\vec R}_n\}$. For $n=1,2$ there exist explicit
formulae:

\begin{eqnarray}
V_1 &=& {\frac{1}{4}} \pi\,d^2 \\
V_2(R/d) &=& {\frac{1}{2}} \pi\,d^2 - {\frac{1}{2}} d^2 \left( \cos^{-1} {R
\over d} - {R \over d}\sqrt{1-{R^2 \over d^2}} \right) \, H(1-R/d)
\label{eq:V1,2}
\end{eqnarray}

\noindent Here $R=|{\vec R}_1-{\vec R}_2|$ and $H(x)$ is the Heavyside step
function. Note that in the scattering context, $d=d(k_z)$, $d$ being the
diameter of the cross-section, which depends on the incident He wavenumber
$k_z$. Using this in Eqs.(\ref{eq:Sn-random}),(\ref{eq:S11})-(\ref{eq:S10}),
and recalling that the present system is translationally invariant, we obtain:

\begin{equation}
S_{00}(R/d) = e^{-\rho\,V_2(R/d)}
\label{eq:S00-random}
\end{equation}

\begin{equation}
S_{11}(R/d) = 1 - 2e^{-{\frac{1}{4}} \rho\,\pi\, d^2} + e^{-\rho\,V_2(R/d)}
\label{eq:S11-random}
\end{equation}

\begin{equation}
S_{01}(R/d) = S_{10}(R/d) = e^{-{\frac{1}{4}} \pi\,\rho\,d^2} -
e^{-\rho\,V_2(R/d)}
\label{eq:S0110}
\end{equation}

\noindent What remains is to insert the last expressions into the general
formula for the scattering intensity, Eq.(\ref{eq:P-S's}), and go through
some algebra and integrations. Defining:

\begin{eqnarray}
\beta = \rho\,d^2 \nonumber\\
\gamma = 2k_z\,h \nonumber\\
\epsilon = {d^2 \over A}
\label{eq:beta-gamma}
\end{eqnarray}

\noindent and:

\begin{equation}
c_1(\beta,\gamma,A) = {(2\pi)^2 \over A} \left(1 + 2e^{-{\frac{1}{4}}
\pi\, \beta} (\cos(\gamma)-1) \right)
\label{eq:c1}
\end{equation}

\begin{equation}
c_2(\beta,\gamma,\epsilon) = 4\pi\epsilon (1-\cos(\gamma))\, e^{-{\frac{1}{2}}
\pi\, \beta}
\label{eq:c2}
\end{equation}

\noindent we obtain:

\begin{equation}
I(\Delta {\vec K}) = {1 \over {(2\pi)^2}} c_1\, \int d{\vec R} \: e^{i {\Delta
{\vec K} \cdot {\vec R}}} + {{e^{{\frac{1}{2}} \pi \, \beta}} \over
{2\pi\,d^2}}\, c_2 \, \int d{\vec R} \: e^{i {\Delta {\vec K} \cdot {\vec R}}}
\,e^{-\rho \,V_2(R/d)}
\label{eq:P-random-1}
\end{equation}

\noindent The first integral evaluates to a delta-function, whereas the second
simplifies to:

\begin{eqnarray}
2\pi e^{-{\frac{1}{2}} \pi \, \beta} \left[ \int_0^d J_0(R\,\Delta K)
e^{{\frac{1}{2}} \beta\, \left( \cos^{-1} {R \over d} - {R \over
d}\sqrt{1-{R^2 \over d^2}} \right)}\, R\,dR + \int_d^\infty J_0(R\,\Delta K)
\, R\,dR \right] \nonumber
\end{eqnarray}

\noindent Let us now define:

\begin{equation}
g(\alpha,\beta) = \int_0^1 J_0(\alpha S) e^{{\frac{1}{2}} \beta \left(
\cos^{-1}S - S \sqrt{1-S^2} \right)} S\,dS ,
\label{eq:g(a,b)}
\end{equation}

\noindent then finally the scattering intensity becomes:

\begin{equation}
I(\Delta {\vec K}) = \delta(\Delta {\vec K}) c_1(\beta,\gamma,A) +
c_2(\beta,\gamma,\epsilon) \left[ g(\Delta K\, d,\beta) + {{J_{-1}(\Delta K\,
d)} \over {\Delta K\, d}}
\right] .
\end{equation}

\newpage

\noindent

\noindent

\noindent

\noindent

\noindent

\noindent

\noindent

\noindent
\begin{figure}
\caption{
A comparison between experimental and theoretical angular intensities for He
scattered from a Pt(111) surface with 50\% adsorbed Ag atoms in different
disorder classes: (a) Translationally random adatoms, (b) Fractal islands, (c)
Size-dispersed compact islands. Insets: Typical configurations in each of the
disorder classes. The orientation of the scattering plane is along the
non-close packed [11\protect\={2}] direction in real space, corresponding to
the close packed $\Gamma_M$ direction in reciprocal space. Thus Bragg peaks
appear at multiples of $4\pi/(\protect\sqrt{3}a)$ where $a=2.77\AA$ is the
Pt(111) lattice constant. Experimental surface temperature: 38K, He beam
energy: 21.6meV ($k_z = 6.43\AA^\protect{-1\protect}$). Solid lines:
theoretical intensities. Circles: experimental data. Note that only the
elastically scattered intensity is shown. The inelastic part was separated in
the experiment using time-of-flight spectroscopy.
}
\label{fig:T-vs-E}
\end{figure}


\newpage


\begin{thebibliography}{10}

\bibitem{1}
{S.C. Wang, G. Ehrlich}, {Surf. Sci.} {\bf 239}, 301 (1990).

\bibitem{2a}
{H. Brune, H. R\"{o}der, C. Boragno and K. Kern}, {Phys. Rev. Lett.} {\bf 73},
1955 (1994).

\bibitem{2b}
{H. R\"{o}der, H. Brune, J.-P. Bucher and K. Kern}, {Surf. Sci.} {\bf 298},
121 (1993).

\bibitem{3}
{P. Zeppenfeld, M. Krzyzowski, Ch. Romainczyk, G. Comsa and M.G. Lagally},
{Phys. Rev. Lett.} {\bf 72}, 2737 (1994).

\bibitem{4}
{G. Rosenfeld, A.F. Becker, B. Poelsema, L.K. Verheij and G. Comsa},
{Phys. Rev. Lett.} {\bf 69}, 917 (1992).

\bibitem{5a}
{R. Stumpf and M. Scheffler}, {Phys. Rev. Lett.} {\bf 72}, 254 (1994).

\bibitem{5b}
{P.Blandin, C. Massobrio and P. Ballone}, {Phys. Rev. B} {\bf 49}, 48 (1994).

\bibitem{6}
{S. Esch, M. Hohage, T. Michely and G. Comsa}, {Phys. Rev. Lett.} {\bf 72},
518 (1994).

\bibitem{7a}
{M. Yanuka, A.T. Yinnon, R.B. Gerber, P. Zeppenfeld, K. Kern, U. Becher and
G. Comsa}, {J. Chem. Phys.} {\bf 99}, 8280 (1993).

\bibitem{7b}
{D.A. Hamburger, A.T. Yinnon, I. Farbman, A. Ben Shaul and R.B. Gerber},
{Surf. Sci.} {\bf 327}, 165 (1995).

\bibitem{7c}
{M. Bott, T. Michely and G. Comsa}, {Surf. Sci.} {\bf 272}, 161 (1992).

\bibitem{8a}
{P. Zeppenfeld, M. Krzyzowski, Ch. Romainczyk, R. David, G. Comsa,
H. R\"{o}der, K. Bromann, H. Brune and K. Kern}, {Surf. Sci. Lett.} {\bf 342},
L1131 (1995).

\bibitem{8b}
{M. Krzyzowski, Ch. Romainczyk, P. Zeppenfeld, R. David and G. Comsa}, {to be
published}.

\bibitem{9}
{A.T. Yinnon, R. Kosloff, R.B. Gerber, B. Poelsema and G. Comsa},
{J. Chem. Phys.} {\bf 88}, 3722 (1988).

\bibitem{9a}
{A.F. Becker, G. Rosenfeld, B. Poelsema and G. Comsa}, {Phys. Rev. Lett.} {\bf
70}, 477 (1993).

\bibitem{9b}
{R. Schinke and R.B. Gerber}, {J. Chem. Phys.} {\bf 82}, 1567 (1985).

\bibitem{9c}
{R.B. Gerber, A.T. Yinnon, M. Yanuka, D. Chase}, {Surf. Sci.}, {\bf 272}, 81
(1992)

\bibitem{17}
{R. B. Gerber}, {Chem. Rev.} {\bf 87}, 29 (1987).

\bibitem{10}
{H. C. Kang and W. H. Weinberg}, {Surf. Sci.} {\bf 299/300}, 755 (1994).

\bibitem{11}
{H. C. Kang and W. H. Weinberg}, {Acc. Chem. Res.} {\bf 25}, 253 (1992).

\bibitem{12}
{C. Uebing and R. Gomer}, {J. Chem. Phys.}, {\bf 95}, 7626 (1991).

\bibitem{13}
{R.B. Gerber, A.T. Yinnon and J.N. Murrell}, {J. Chem. Phys.}, {\bf 31}, 1
(1978).

\bibitem{14}
{A.T. Yinnon, R.B. Gerber, D.K. Dacol and H. Rabitz}, {J. Chem. Phys.} {\bf
84}, 5955 (1986).

\bibitem{15}
{D.K. Dacol, H. Rabitz and R.B. Gerber}, {J. Chem. Phys.} {\bf 86}, 1616
(1987).

\bibitem{16}
{R.B. Gerber, Delgado-Bario, editor, in {\em Dynamics of Molecular Processes}},
{p.299}, {(IOP, Bristol, 1993)}.

\bibitem{19}
{B. Poelsema and G. Comsa, in {\em Scattering of Thermal Energy Atoms from
Disordered Surfaces}, Springer Tracts in Modern Physics, Vol. 115 (Springer,
Berlin, 1989).}

\bibitem{18}
{A.T. Yinnon, R. Kosloff and R.B. Gerber}, {J. Chem. Phys.} {\bf 88}, 7209
(1988).

\bibitem{19b}
{D.A. Hamburger and R.B. Gerber}, {J. Chem. Phys.} {\bf 102}, 6919 (1995).

\bibitem{19a}
{In fact the potential was obtained for He/Pt(110), but should likewise be
applicable to the Pt(111) face since the potential is a {\em laterally
averaged} one.}

\bibitem{20}
{M.A. Krzyzowski, P. Zeppenfeld and G. Comsa}, {J. Chem. Phys.} {\bf 103},
8705 (1995).


\bibitem{21}
{A.M. Lahee, J.P. Manson, J.P. Toennies and C. W\"{o}ll}, {J. Chem. Phys.} {\bf
86}, 7194 (1987).

\bibitem{20a}
{N. Esbjerg and J.K. Norskov}, {Phys. Rev. Lett.} {\bf 45}, 807 (1980);
{N.D. Lang and J.K. Norskov}, {Phys. Rev. B} {\bf 27}, 4612 (1983);
{J.K. Norskov, K.W. Norskov, P. Stoltze and L.B. Hansen}, {Surf. Sci.} {\bf
283}, 277 (1993).

\bibitem{21aaa}
{M. Bott, M. Hohage, M. Morgenstern, T. Michely and George Comsa},
{Phys. Rev. Lett.} {\bf 76}, 1304 (1996)

\bibitem{21aa}
{H. Brune, C. Romainczyk, H. R\"{o}der, and K. Kern}, {Nature} {\bf 369}, 469
(1994).

\bibitem{21a}
{D.A. Hamburger, A.T. Yinnon and R.B. Gerber}, {Chem. Phys. Lett.} {\bf 253},
223 (1996)

\bibitem{8}
{P.Pfeifer and M.W. Cole}, {New J. of Chem.} {\bf 14}, 221 (1990).

\bibitem{22}
{D.A. Hamburger-Lidar}, {Phys. Rev. E.}, {\bf 54}, 354 (1996).

\bibitem{comment}
{It may be appropriate at this point to explain why we did not use the KMC
method 
to generate configurations at 38K to guide us in the search for a model for
the surface morphology. The reason is that, as explained in Sec.\ref{KMC}, we
do not have an absolute temperature scale available in the KMC
simulations.}

\bibitem{23}
{M.A. Krzyzowski, P. Zeppenfeld and G. Comsa}, {to be published.}

\bibitem{Roder}
{H. R\"{o}der, thesis Nr. 1288, EPF Lausanne (1994)}.

\bibitem{25}
{S. Torquato and G. Stell}, {J. Chem. Phys.} {\bf 77}, 2071 (1982); {\bf 78},
3262 (1983); {\bf 79}, 1505 (1983).


\end{thebibliography}
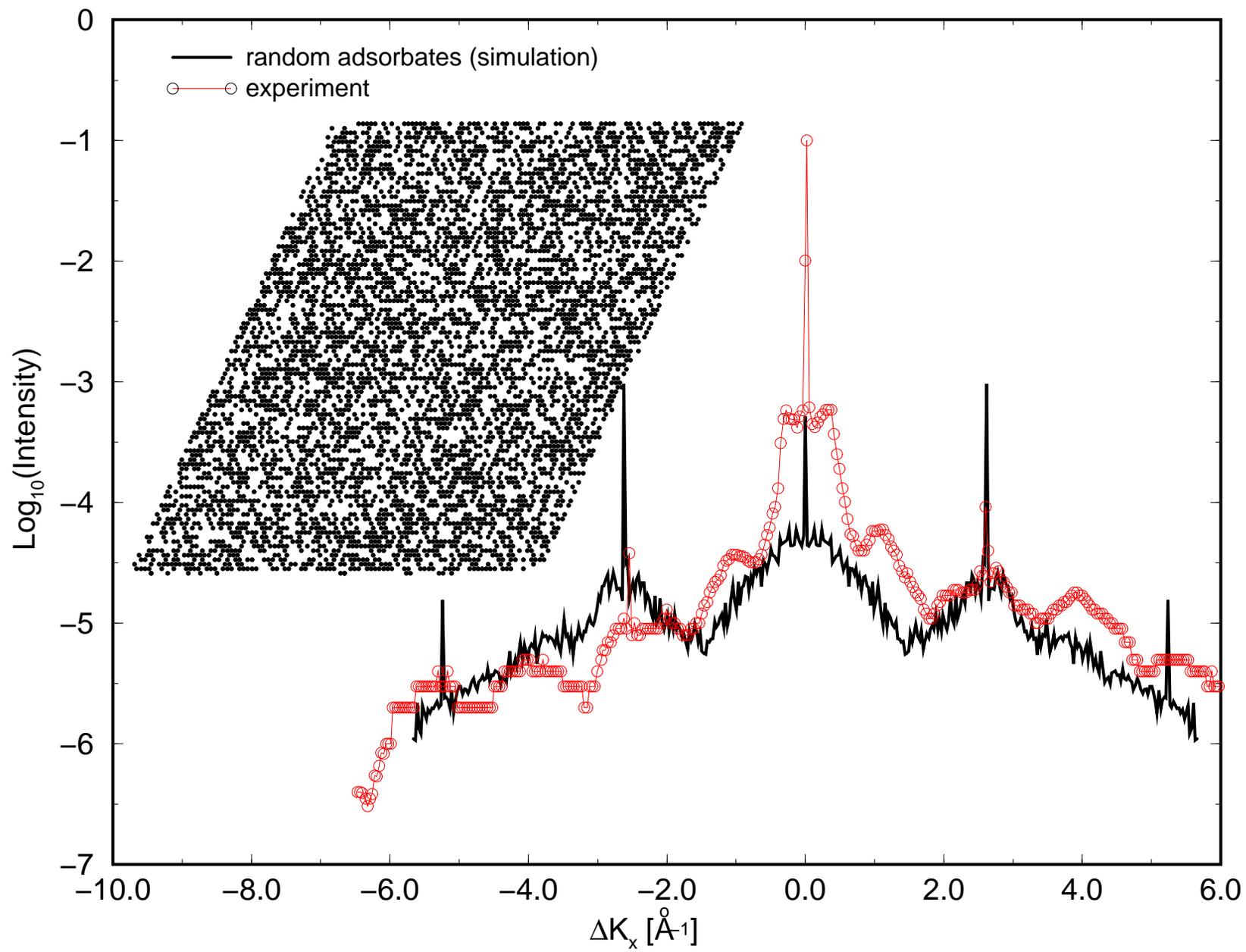
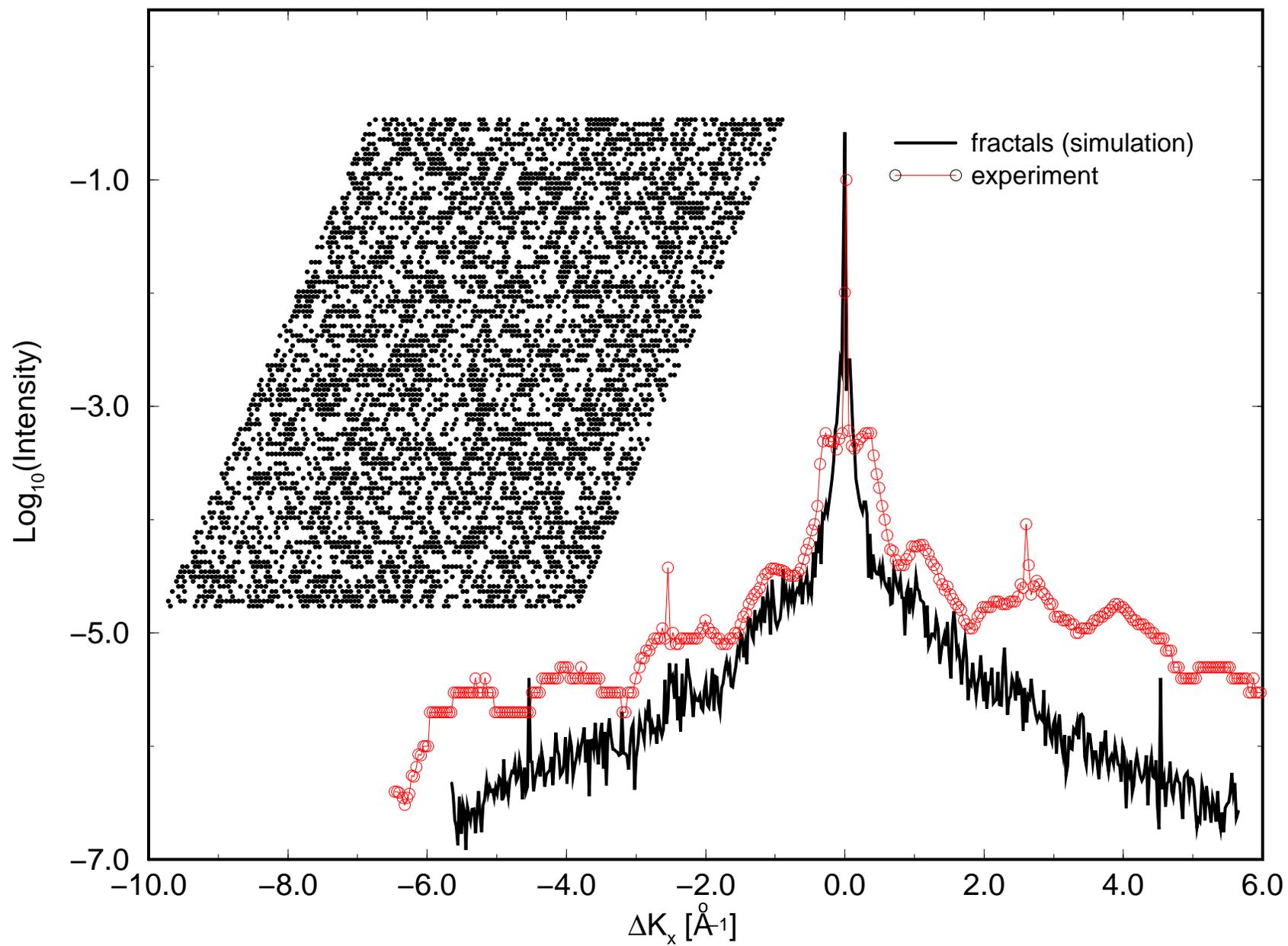
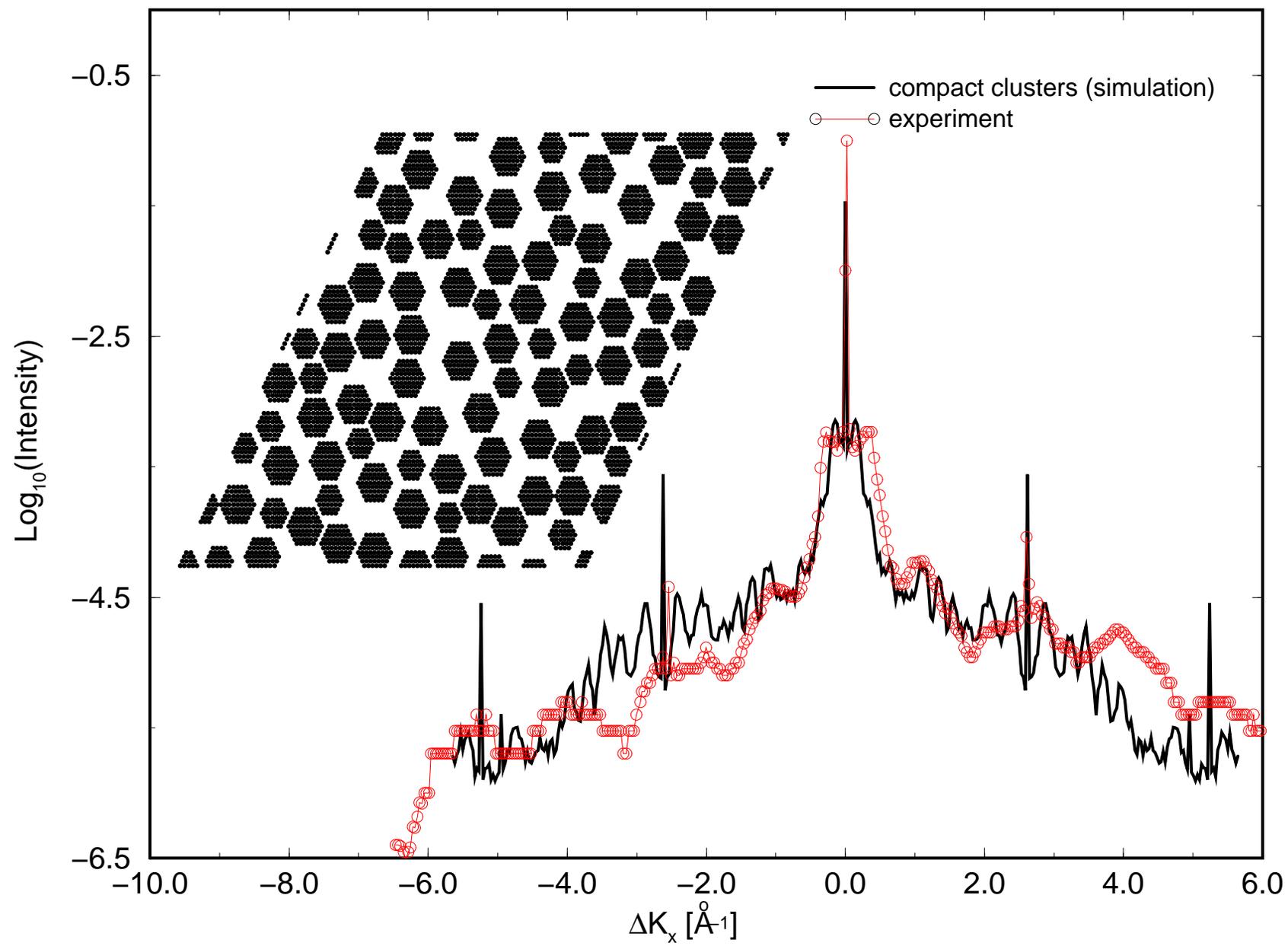
\end{document}